%% file: main.tex
\documentclass[11pt]{article}
\usepackage{graphicx}
\input{preamble}
\usepackage{fancyhdr}

\def\manuscriptformat{arxiv}
\def\biostatisticsformat{biostatistics}
\newif\ifbiostatistics
\ifx\manuscriptformat\biostatisticsformat
\biostatisticstrue
\else
\biostatisticsfalse
\fi

\newcommand{\runningtitle}{Characterizing Survivor Principal Strata Over Time}

\fancypagestyle{biostatistics}{
\fancyhf{}
\fancyhead[C]{\small\itshape\runningtitle}
\fancyfoot[C]{\thepage}

}
\newcommand{\biostatisticscoverinfo}{
\vspace{2em}

\noindent\textbf{\textsuperscript{\(\dagger\)}Corresponding author.}
Keith Barnatchez, Department of Biostatistics, Johns Hopkins Bloomberg
School of Public Health, 615 N Wolfe St, Baltimore, MD 21205, USA.
Email: \href{mailto:kbarnat1@jh.edu}{kbarnat1@jh.edu}

\vspace{1em}

\noindent\textbf{Funding.}
This work was supported in part by National Institutes of Health Grant
R01DA056407.

\vspace{1em}

\noindent\textbf{Conflict of Interest.}
W.B. and J.G. are current and former employees of the Neurological Clinical Research Institute (NCRI) at Mass General Brigham, respectively. K.B. previously received doctoral funding support from the NCRI. The NCRI manages the PRO-ACT database used in this study but had no role in the conduct or reporting of the work. The authors declare no other conflicts of interest relevant to this work.

\vspace{1em}

\noindent\textbf{Data Availability.}
The data analyzed in this study were obtained from the PRO-ACT database. Patient-level PRO-ACT data are available to registered users under the PRO-ACT Terms and Conditions and cannot be redistributed by the authors. Aggregate results and analysis code may be shared in accordance with those terms. Access to the database may be requested at \url{https://ncri1.partners.org/ProACT/}. Code to replicate the simulation exercises is available at
\url{https://github.com/keithbarnatchez/survstrat-paper}, and an R package implementing the proposed methods is available at
\url{https://github.com/keithbarnatchez/survstrat}.

\vspace{1em}

\noindent\textbf{Acknowledgments.}
We thank Sophie Woodward and Eric Macklin for helpful comments and suggestions. Large language models (Anthropic's Claude) were used to assist with code development and review, and to identify typos and inconsistencies in the manuscript. All methodological development, analyses, and conclusions are the authors' own, and the authors take full responsibility for the content of the manuscript.
}

\title{
\ifbiostatistics\vspace{-2.5em}\else\vspace{-3.5em}\fi
Characterizing Survivor Principal Strata Over Time:
Identification, Efficient Estimation, and Sensitivity Analysis
\\[1em]
}

\author{
Keith Barnatchez\ifbiostatistics$^{1,\dagger}$\else$^{1}$\fi,
Willow Butler$^{2}$, Julia A. Geller$^{2,3}$, Elizabeth A. Stuart$^{1}$, \\[0.5em]
{\small The Pooled Resource Open-Access ALS Clinical Trials Consortium$^*$} \\[1.2em]
\small
$^{1}$Department of Biostatistics, Johns Hopkins Bloomberg School of Public Health \\
\small $^{2}$Neurological Clinical Research Institute, Massachusetts General Hospital \\
\small $^{3}$Computer Science and Engineering Division, University of Michigan
}

\begin{document}

\maketitle
  \thispagestyle{empty}

  \ifbiostatistics
    \biostatisticscoverinfo
    \clearpage
  \fi

  \setlength{\parskip}{0pt}

\begin{abstract}
Longitudinal studies with functional outcomes subject to truncation-by-death frequently report time-indexed survivor average causal effect estimates, defined among individuals who would survive regardless of treatment assignment.  Because the survivor stratum can change over time, the resulting effect estimates may pertain to populations with meaningfully different characteristics at different times. To better understand these time-varying strata, we propose accompanying survivor average causal effect estimates with measures of covariate divergence between the survivor and baseline populations. We study covariate-specific standardized mean differences (SMDs) and the Kullback–Leibler (KL) divergence as examples of moment-based and distributional summaries, respectively. We show that these divergence measures are identified under a subset of standard principal stratification assumptions and that their identification involves nuisance functions that are already routinely estimated in survivor analyses, meaning the proposed diagnostics can be computed at little additional cost. We derive the efficient influence functions for these quantities, enabling the construction of one-step bias-corrected estimators with appealing large-sample properties and allowing the use of machine learning models for nuisance function estimation. We demonstrate that in randomized studies, one-step estimation of  SMDs is robust to arbitrary nuisance model misspecification, whereas the KL divergence and broader density ratio estimands are more sensitive to nuisance misspecification.  Additionally, we provide a sensitivity analysis framework that partially relaxes a monotonicity assumption required for identification, and construct one-step estimators for the bounds obtained through this framework. Through simulation studies, we demonstrate the finite-sample performance of the proposed estimators and illustrate how reporting divergence measures alongside causal effect estimates can improve the interpretability of survivor analyses. Finally, we assess covariate divergence in survivor strata over time in amyotrophic lateral sclerosis (ALS) trials, using data from the Pooled Resource Open-Access ALS Clinical Trials database.
\end{abstract}
\textit{Key words:} Amyotrophic lateral sclerosis, causal inference, covariate divergence, influence function, principal stratification, sensitivity analysis

\ifbiostatistics
    \thispagestyle{empty}
    \clearpage
    \setcounter{page}{1}
    \pagestyle{biostatistics}
  \else
    \setcounter{page}{0}
    \clearpage
  \fi

\ifbiostatistics\else\doublespacing\fi

\section{Introduction}
\label{sec:introduction}

Biomedical studies often aim to estimate the effect of a baseline treatment on outcomes measured repeatedly over a follow-up period. In many applications, it is common for individuals to not survive the full study duration. For functional, non-survival outcomes, death complicates causal inference since  outcomes are not well-defined after death \citep{zhang2003estimation}. This problem---commonly referred to as \textit{truncation by death}---has motivated a large literature on causal estimands and estimation strategies \citep{luo2023causal, li2025causal}.

A common approach to address truncation by death is to restrict focus to identifying and estimating causal effects within the principal stratum of individuals who would survive under either treatment level, commonly referred to as the \textit{survivor} principal stratum \citep{rubin2006causal}. Membership in the survivor principal stratum is latent because each participant is observed under only one treatment level, whereas determining membership requires knowing whether that participant would have survived under both treatment levels. At a given follow-up time, the corresponding survivor average causal effect (SACE) compares potential outcomes among individuals who %would survive to that time point under either treatment condition. 
are members of the latent stratum of survivors, since restricting attention to this stratum avoids causal contrasts involving outcomes that would be undefined under one treatment condition \citep{grossi2025bayesian}. This estimand has been studied extensively in the principal stratification and censoring due to death literatures \citep{tchetgen2014identification}, where these fields have seen extensive methods development for the identification and estimation of survivor average causal effects \citep{wang2017identification,colantuoni2018statistical, zehavi2023matching,lu2026principal,grossi2025bayesian},  with recent work considering effect modification \citep{gonccalves2025effect}, policy learning \citep{chu2023multiply,park2025evaluating}, and sensitivity analyses \citep{tong2025semiparametric,tong2025doubly,imai2008sharp}.

Despite these methodological developments, there are concerns about the interpretability and scientific utility of survivor average causal effects, especially in clinical trials where the handling of post-treatment intercurrent events has experienced increased regulatory attention \citep{pearl2011principal,ich2019addendum,fleming2025perspective,vansteelandt2025chasing,ding2025roles}. In addition to requiring strong identifying assumptions, the latent survivor principal stratum  may differ substantially from the population enrolled at baseline. Even when a survivor average causal effect is well-defined and identified, it may pertain to a subgroup whose  characteristics are not especially representative of the baseline trial population. In turn, its relevance for a clinical population of interest depends in part on how similar the survivor stratum is to the population from which study participants were drawn, a concern that is central to the literature on generalizability and transportability \citep{degtiar2023review}. This concern is especially pronounced in longitudinal studies, since the survivor stratum can itself shift across follow-up, so that the population indexed by a survivor average causal effect need not remain fixed over time.

Summarizing a latent principal stratum is closely related to profiling compliers in instrumental variable analyses, where the complier covariate distribution is not directly observable \citep{angrist2010extrapolate}. In these settings, \citet{abadie2003semiparametric} showed that moments of complier characteristics can be represented as weighted moments of the observed data, and \citet{singh2024double} developed doubly robust, semiparametric efficient estimators for the same class of parameters. Related work has used covariate-indexed summaries to describe how compliers differ from the full sample \citep{marbach2020profiling}. These methods, however, were developed (i) for the setting of handling noncompliance rather than truncation due to death, and (ii) for cross-sectional rather than longitudinal settings.  In truncation by death settings, the latent principal stratum is defined through survival under both treatment conditions and is indexed by follow-up time, so that its composition can shift as the surviving population changes. For this reason, it can be useful to report measures quantifying how the survivor stratum differs from the baseline population at each follow-up time. These summaries can be especially informative in randomized trials, where truncation by death is common and treatment is assigned by a known mechanism. The known treatment mechanism also has implications for estimation and inference that complier-profiling work, primarily developed for observational settings, has not examined. Moreover, covariate-specific summaries become impractical to inspect across many covariates and follow-up times, suggesting alternative scalar-level divergence summaries can serve as useful complements to these measures.

\paragraph{Contributions.} In this paper, we assess how the composition of survivor principal strata changes over follow-up  using time-indexed \textit{divergence measures} that compare the corresponding survivor covariate distribution with the distribution among all enrolled participants. We study two  classes of divergence measures: standardized mean differences (SMDs) and the Kullback-Leibler (KL) divergence. SMDs are time-indexed analogues of the complier means studied in the instrumental variables literature \citep{abadie2003semiparametric,singh2024double}, whereas the KL divergence is a scalar summary capable of capturing higher-order discrepancies between the joint baseline and survivor covariate distributions.

Our work makes three main contributions. \textit{First}, we demonstrate that these measures are identified under a subset of the assumptions typically used to identify survivor average causal effects. In particular, their identification does not require the cross-world conditions needed for the SACE estimands themselves. The resulting identifying functionals depend only on the conditional survival probability under control, a nuisance function that is already estimated in standard survivor analyses. 
\textit{Second}, we derive efficient influence functions for these identifying functionals, and use them to construct efficient one-step estimators that allow flexible estimation of the nuisance functions while supporting valid asymptotic inference. We further establish an important robustness asymmetry between the two summaries: in randomized trials, where the propensity score is known by design, one-step SMD estimators remain consistent and asymptotically normal regardless of how  survival  is modeled, whereas one-step KL divergence estimators, along with estimators of the broader class of density-ratio divergences, require correct specification of the survival model. In this sense, the proposed moment-based summaries are robust to misspecification of the survival mechanism, while the density-ratio divergences are more sensitive to higher-order distributional differences, but are more reliant on correct nuisance specification.
\textit{Finally}, we  develop a sensitivity analysis framework that partially relaxes the monotonicity assumption used for identification. Rather than requiring the analyst to specify an entire covariate-indexed sensitivity function, our proposed sensitivity framework is carried out over a grid of sensitivity parameters characterizing the mean rate of monotonicity violations and the extent to which those violations vary across covariate strata. 

\paragraph{Paper organization.} The remainder of this paper is structured as follows. Section \ref{sec:prob-setting} introduces the problem setting of interest and our proposed divergence measures, while Section \ref{sec:identification} outlines conditions that enable their identification from observed data. We provide plug-in and debiased one-step estimators of these measures in Section \ref{sec:methods}, characterizing conditions that enable valid asymptotic inference. While our methods can accommodate both observational and randomized studies, we additionally study the asymptotic behavior of our methods in the special case where the data arise from a randomized trial. Section \ref{sec:relaxing-monotonicity} presents a sensitivity analysis to partially relax an untestable monotonicity condition our methods rely on for identification. We study the finite-sample performance of our proposed methods and sensitivity analysis framework in Section \ref{sec:simulation},  additionally demonstrating informative approaches to report our proposed measures when estimating SACEs. In Section \ref{sec:application}, we apply these methods to data from the Pooled Resource Open-Access ALS Clinical Trials (PRO-ACT) database, a publicly available repository that aggregates individual-participant data from over forty industry- and academic-sponsored amyotrophic lateral sclerosis phase II/III clinical trials. Finally, Section \ref{sec:discussion} concludes with a discussion and considerations for future work.

\section{Problem Setting}
\label{sec:prob-setting}

\subsection{Data Structure}
We consider a hypothetical longitudinal study, where interest lies in the effect of a binary treatment, administered at baseline, on a time-varying outcome of interest as well as survival. We suppose the researcher observes $n$ independent copies of the observational unit
\[
\bm O = (S_0\cdot Y_0,\  S_0,\ \ldots, \  S_T\cdot Y_T, \ S_T, \ A, \ \bX), \
\]
where $A$ is a binary treatment indicator,  $\bX$ is a $p$-dimensional vector of baseline covariates, $S_t$ indicates whether the participant survived up to time $t$, and $Y_t$ is the outcome of interest at time $t$ that is measured if the participant survived to time $t$, which we refer to as the ``functional" outcome as it often measures patient functioning. Accordingly, $Y_t$ is undefined when $S_t=0$ but defined and measured otherwise. To capture this, we follow \cite{rubin2006causal} and define $S_t\cdot Y_t = Y_t$ when $S_t=1$, and $S_t \cdot Y_t = *$ when $S_t=0$. Throughout, we adopt the Rubin potential outcomes framework \citep{rubin1974estimating}, letting $S_t(a)$ and $Y_t(a)$ denote the potential survival status and functional outcome under treatment $A=a$, respectively. 

Here and throughout, a participant is considered a time-$t$ survivor if $S_t(1)=S_t(0)=1$. Membership in this subgroup is latent, since survival status is only observed under one treatment level. More broadly,  the pair $(S_t(1),S_t(0))$ defines four latent subgroups, which group participants according to how they would survive under both treatment levels, regardless of the treatment they were actually assigned. Subgroups that themselves are defined by potential outcomes are typically referred to as \textit{principal strata}, and the broader field of identifying and estimating causal effects among principal strata is commonly referred to as \textit{principal stratification} \citep{frangakis2002principal}.

A key causal quantity of interest in studies with this form is the time-indexed survivor average causal effect (SACE) $\tau_t = \E[Y_t(1) - Y_t(0) \mid S_t(0)=S_t(1)=1].$
When interested in this contrast, restriction to survivors is necessary because average causal contrasts are undefined for subsets of individuals where $Y_t(1)$ or $Y_t(0)$ are undefined \citep{rubin2006causal}. While $\tau_t$ is a standard estimand in settings with censoring due to death, with numerous works centered around its identification and efficient estimation  \citep{wang2017identification,tchetgen2014identification,zehavi2023matching},
the baseline characteristics of survivors at any time point $t$ may be substantially different than the general baseline covariate distribution. In this sense, survivors may be substantially different than the baseline clinical population of interest, and in fact differences should be expected when baseline characteristics are prognostic for survival.  Further, the set of survivors for whom $S_t(0)=S_t(1)=1$ itself can vary in $t$, implying $\tau_t$---despite being well-defined at all $t$---corresponds to a different set of participants at each time point $t$. To aid researchers in the interpretation of $\tau_t$, we propose accompanying estimates of $\tau_t$ with summary measures that quantify how the time-$t$ survivor covariate distribution differs from the baseline covariate distribution. 

\subsection{Target Parameters}

In this Section, we present target parameters that summarize  how the latent time-$t$ survivor covariate distribution differs from the baseline covariate distribution among all participants at enrollment. In practice, these measures can be reported alongside estimates of $\tau_t$ to quantify how similar the time-$t$ survivor stratum is to the baseline study population, which we demonstrate in Sections \ref{sec:simulation} and \ref{sec:application}. For brevity, we consider two divergence measures throughout: (i) SMDs, and (ii) KL divergence.  We begin by considering SMDs of the form
\[
\delta_{kt} = (\E[X_k] - \E[X_{k} \mid S_t(0)=S_t(1)=1])/\text{SD}(X_k),
\]
where $\bX = (X_1,\ldots,X_p)$ and we let $\bm \delta_t := (\delta_{1t},\ldots,\delta_{pt})$ collect all covariate-indexed SMDs. $\delta_{kt}$ captures the expected difference between the time-$t$ survivor and baseline distribution for the $k$-th covariate $X_k$, normalized to the scale of the baseline $X_k$ distribution. Particularly in biomedical applications, SMDs are a widely-used metric  for quantifying similarities between covariate distributions across groups of interest, and  are commonly used as a diagnostic for assessing covariate balance across treatment categories in the matching and weighting literature \citep{austin2009balance,stuart2010matching} that have been used to characterize compliers in instrumental variable applications \citep{abadie2003semiparametric,singh2024double}. 

Alongside $\delta_{kt}$, we additionally consider the time-$t$ \textit{KL divergence}:
\[
\psi_t = \E\left[\log \left(\frac{p(\bX|S_t(0)=S_t(1)=1)}{p(\bX)} \right) \Bigg| \ S_t(0)=S_t(1)=1 \right].
\]
Heuristically, $\psi_t$ quantifies the degree of information loss from attempting to approximate the time-$t$ survivor covariate distribution $p(\bX|S_t(0)=S_t(1)=1)$ with the baseline covariate distribution $p(\bX)$. 
Recalling that KL divergence is asymmetric in its arguments, we choose the above formulation over its flipped counterpart $\E[\log\{p(\bX)/p(\bX|S_t(0)=S_t(1)=1)\}]$ because $\psi_t$ remains bounded as $p(\bX | S_t(0)=S_t(1)=1) \rightarrow 0$, whereas estimators of the  flipped functional can be particularly sensitive to strata of $\bX$ where $p(\bX | S_t(0)=S_t(1)=1)$ is near-zero.
KL divergence is a standard metric for comparing the similarity of covariate distributions in the broader statistics literature, with recent applications in causal inference as a matching diagnostic \citep{yu2021information}. Additionally, entropy balancing approaches in the  covariate balancing literature minimize a KL-type metric across groups subject to covariate balance constraints \citep{zhao2017entropy,josey2021transporting}. Although slightly less straightforward to interpret, $\psi_t$ can capture more nuanced, higher-order distributional differences, whereas  $\bm \delta_t$ summarizes only first-moment discrepancies between survivor and baseline covariate distributions.
In this sense, the two summaries capture complementary features of the survivor covariate distribution. In Section \ref{sec:methods}, we demonstrate that $\psi_t$'s ability to capture higher-order discrepancies comes at the cost of requiring stronger conditions on nuisance function estimation relative to estimation of $\delta_{kt}$. 

More generally, $\psi_t$ is one of many scalar \emph{distributional} divergences $\E[h(w_0(\bX,t))]$, indexed by a transformation $h$ of the survivor-to-baseline density ratio. This broad set of measures, which we further discuss in the Supplementary Material, includes  the chi-square and Hellinger distances. For brevity, however, we focus  on $\bm \delta_t$ and $\psi_t$ throughout.

\subsection{Assumptions}

We now discuss conditions that enable the above divergence measures to be estimated from the observed data structure $\bm O$. The identification of $\bm \delta_t$ and $\psi_t$ requires assumptions commonly invoked in the principal stratification literature:
\begin{assumption}[Consistency]
\label{as:consistency}
  $S_t = A\cdot S_t(1) + (1-A) \cdot S_t(0)$ for all $t$
\end{assumption}
\begin{assumption}[Positivity]
\label{as:positivity}
  $\PP(A=1|\bX) \in (0,1)$  
\end{assumption} 
\begin{assumption}[Survival Unconfoundedness]
\label{as:surv-unc}
    $S_t(a) \indep A \mid \bX$ for all $t$ and $a$
\end{assumption}
\begin{assumption}[Monotonicity]
\label{as:monotonicity}
    $\PP(S_t(1) \geq S_t(0))=1$ for all $t$
\end{assumption}
Above, Assumptions \ref{as:consistency}-\ref{as:surv-unc} are standard conditions that tend to hold by design in randomized trials and are commonly invoked in the analysis of non-randomized studies, where Assumption \ref{as:surv-unc} functions as a standard no unmeasured confounding condition.  Assumption \ref{as:monotonicity} is a strong condition that implies treatment cannot cause death for any individual. The validity of Assumption \ref{as:monotonicity} will vary across specific applications, and should ideally be assessed in consultation with subject-matter experts for any particular application. Numerous sensitivity analyses have been proposed that partially relax this condition \citep{shepherd2011sensitivity,ding2011identifiability,tong2025doubly} when interested in the SACE $\tau_t$, and we consider similar sensitivity analyses for $\bm \delta_t$ and $\psi_t$ in Section \ref{sec:relaxing-monotonicity}.

In practice, $\bm \delta_t$ and $\psi_t$ are  primarily useful when they accompany estimates of the time-$t$ survivor average causal effect $\tau_t$. Though this paper  focuses on estimation of $\bm \delta_t$ and $\psi_t$, for completeness, we  present additional assumptions needed to identify $\tau_t$, but not needed to identify $\bm \delta_t$ or $\psi_t$:
\begin{assumption}[Primary Outcome Unconfoundedness and Consistency]
\label{as:out-unc}
$Y_t(a) \indep A \mid \bX, S_t(a)=1$ for all $a,t$, and $A=a, S_t=1 \implies $ $Y_t(a)=Y_t$ for all $t \in \{0,\ldots,T\}$ 
\end{assumption}
\begin{assumption}[Cross-world survival mean exchangeability]
\label{as:cross-ex}
$\E[Y_t(1)|S_t(1)=1, S_t(0)=1, \bX] = \E[Y_t(1)|S_t(1)=1, \bX]$ for all $t \in \{0,\ldots,T\}$    
\end{assumption}
Assumptions \ref{as:out-unc} and \ref{as:cross-ex}, and closely related variants, are commonly invoked in the truncation-by-death literature \citep{tchetgen2014identification} and the broader principal stratification literature. While Assumption \ref{as:out-unc} is a relatively mild condition, Assumption \ref{as:cross-ex} is a  strong condition that asserts a cross-world conditional mean independence between the functional outcome under treatment, $Y(1)$, and survival under the control condition, $S(0)$. Practically, Assumption \ref{as:cross-ex} requires that knowledge of survival status under the control condition provides no additional information about potential outcomes under treatment, after conditioning on survival under treatment and any baseline covariates. 
Like the monotonicity Assumption \ref{as:monotonicity}, Assumption \ref{as:cross-ex} is a strong requirement whose validity should be assessed in consultation with subject matter experts, and provides a design-based rationale to collect as many baseline prognostic variables as possible \citep{miles2017partial}. Importantly, the identifying functionals for $\bm \delta_t$ and $\psi_t$ will hold under Assumptions \ref{as:consistency}-\ref{as:monotonicity}, regardless of whether Assumptions \ref{as:out-unc} or \ref{as:cross-ex} hold.

\section{Identification and Efficiency Theory}
\label{sec:identification}
In this Section, we provide identifying functionals for both $\bm \delta_t$ and $\psi_t$, as well as their corresponding efficient influence functions (EIFs). Throughout, our identification and efficiency results are pointwise in $t$ to enable fully nonparametric inference. We discuss approaches that impose longitudinal structure on $g_0(\bX,t)$ in Sections \ref{sec:rand-trial} and \ref{sec:discussion}, and additionally present identifying functionals that accommodate right censoring due to dropout in the Supplementary Material, focusing on the present setting with no dropout for simplicity. 

\begin{table}[t!]
    \centering
    \begin{tabular}{lll}
    \toprule
    Nuisance & Definition & Interpretation \\ \midrule
        $g_{a}(\bX,t)$ & $\PP(S_t=1|A=a,\bX)$ & Time-$t$ conditional survival probability \\
        $e(\bX)$ & $\PP(A=1|\bX)$ & Treatment propensity score \\
        $w_{a}(\bX,t)$ & $g_{a}(\bX,t)/\mu_{at}$ & Normalized time-$t$ survival probability \\ 
        $\mu_{at}$ & $\E[g_{a}(\bX,t)]$ & Marginal time-$t$ survival probability \\
        % \midrule
        % $\theta_{kt}$ & $\E[X_k \mid S_t(0)=S_t(1)=1]$ & Always-survivor covariate mean \\
        % $\psi_t$ &  & KL divergence 
        \bottomrule
    \end{tabular}
    \caption{Notation for nuisance functions and key quantities.}
    \label{tab:nuisances}
\end{table}

Table \ref{tab:nuisances} provides definitions for nuisance functions and quantities referenced throughout.
We first present an identifying functional for $\theta_{kt} := \E[X_k|S_t(0)=S_t(1)=1]$, noting $\theta_{kt}$ is the only unidentified component of the SMD $\delta_{kt} = \{\E[X_k] - \theta_{kt} \}/\text{SD}(X_k)$:
\begin{theorem}
\label{thm:theta-id}
Under Assumptions \ref{as:consistency} through \ref{as:monotonicity}, the survivor covariate mean $\theta_{kt}$ is identified by
\begin{equation}
    \theta_{kt} =  \frac{\E[X_k \cdot g_0(\bX,t)]}{\E[g_0(\bX,t)]} = \E[X_k \cdot w_0(\bX,t) ].
    \label{eq:theta-id}
\end{equation}
\end{theorem}
Under Assumption \ref{as:monotonicity}, survival under control implies survival under treatment, meaning $g_0(\bX,t)$ can be viewed as the conditional probability of time-$t$ survivor principal strata membership. Intuitively, \eqref{eq:theta-id} implies that one can reweight the baseline covariate distribution of $X_k$ to mimic that of the time-$t$ survivor covariate strata through the normalized survivor strata membership weights $w_0(\bX,t)$. This functional is analogous to the identifying functionals discussed in \cite{abadie2003semiparametric} and \cite{singh2024double}, who focused on summarizing the covariate distribution of the complier principal stratum  in instrumental variables settings. Here, the weights $w_0(\bX,t)$ are additionally indexed by time. An immediate corollary of Theorem \ref{thm:theta-id} is that the covariate-level SMD $\delta_{kt}$ is identified by
\begin{equation}
    \label{eq:delta-id}
    \delta_{kt} = \left(\E[X_k] - \frac{\E[X_k \cdot g_0(\bX,t)]}{\E[g_0(\bX,t)]}\right) \cdot \frac{1}{\text{SD}(X_k)},
\end{equation}
where one can collect each identified component in \eqref{eq:delta-id} to identify $\bm \delta_t$. 
An identifying functional for the KL divergence $\psi_t$ is presented in Theorem \ref{thm:kl-id} below.
\begin{theorem}
    \label{thm:kl-id}
    Under Assumptions \ref{as:consistency} through \ref{as:monotonicity}, $\psi_t$ is identified by
    \begin{equation}
        \label{eq:psi-id}
           \psi_t = \E[w_{0}(\bX,t) \log w_{0}(\bX,t)].
    \end{equation}
\end{theorem}
Under Assumptions \ref{as:consistency}-\ref{as:monotonicity} and analogous to $\theta_{kt}$, $\psi_t$ can be written as a function of the principal scores $g_0(\bX,t)$. Intuitively, notice  $\psi_t=0$ when $w_{0}(\bX,t)=1$, which occurs when there is no shift between the baseline and survivor covariate distributions. Similarly, when $g_0(\bX,t)$ varies strongly in $\bX$, the ratio $w_0(\bX,t)$ will tend to be large, consistent with large values of $\psi_t$. Notably, $\psi_t$ is a nonlinear functional of the survival regression $g_0(\bX,t)$, whereas $\theta_{kt}$ is the ratio of two linear functionals. This nonlinearity presents challenges for estimating $\psi_t$, which we discuss further in Section \ref{sec:methods}.

We now present the efficient influence functions (EIFs) for $\theta_{kt}$ and $\psi_t$. EIFs characterize the semiparametric efficiency bound for any regular and asymptotically linear (RAL) estimator of a statistical quantity, and are a crucial ingredient for constructing debiased estimators that enable researchers to conduct valid inference when estimating nuisance functions with flexible machine learning methods \citep{kennedy2024semiparametric} that possess slow convergence rates.  The EIFs for $\theta_{kt}$ and $\psi_t$, which will be used to construct debiased estimators in Section \ref{sec:methods}, are provided in Theorem \ref{thm:eif-thm} below:
\begin{theorem}
\label{thm:eif-thm}
The EIFs for $\theta_{kt}$ and $\psi_t$ in a nonparametric model for the observed data are
\begin{align*}
 & \varphi_{\theta_{kt}}(\bm O)
=
\frac{1}{\mu_{0t}}
\left[
\frac{\mathbb{I}(A=0)}{1-e(\bX)} \{S_t - g_0(\bX,t)\}(X_k - \theta_{k,t})
+
g_0(\bX,t)(X_k - \theta_{k,t})
\right],   \ \ \ \text{and} \\
&\varphi_{\psi_t}(\bm O) = 
\frac{\mathbb{I}(A=0)}{1-e(\bX)}
\frac{\log w_{0}(\bX,t)-\psi_t}{\mu_{0t}}
\bigl(S_t-g_0(\bX,t)\bigr)
+
w_0(\bX,t)\bigl(\log w_0(\bX,t)-\psi_t-1\bigr)
+1.
\end{align*}
\end{theorem}
Both $\varphi_{\theta_{kt}}(\bm O)$ and $\varphi_{\psi_{t}}(\bm O)$ contain two nuisance functions: the conditional survivor probability $g_0(\bX,t)$ appearing in the identifying functionals \eqref{eq:theta-id} and \eqref{eq:psi-id}, and additionally the treatment propensity score function $e(\bX)$. One-step estimators for the SACE similarly depend on both $g_0(\bX,t)$ and $e(\bX)$, meaning the estimators proposed in Section \ref{sec:methods} can be incorporated into SACE analyses using the same nuisance estimates.
Noting $\theta_{kt} = N_{kt}/D_t$ with $N_{kt} = \E[X_k \cdot g_0(\bX,t)]$ and $D_t = \mu_{0t}$,  $\varphi_{\theta_{kt}}(\bm O)$ can alternatively be expressed through the EIFs of the $N_{kt}$ and $D_t$. Specifically, in the Supplementary Material we show
\[
\varphi_{N_{kt}}(\bm O) = \frac{\II(A=0)}{1-e(\bX)}\{S_t - g_0(\bX,t)\}X_k + g_0(\bX,t) X_k - N_{kt},
\]
\[
\varphi_{D_t}(\bm O) = \frac{\II(A=0)}{1-e(\bX)}\{S_t - g_0(\bX,t)\} + g_0(\bX,t) - D_t,
\]
where one can then recover $\varphi_{\theta_{kt}}(\bm O)$ by noting $
\varphi_{\theta_{kt}}(\bm O) = \frac{1}{D_t}\left[\varphi_{N_{kt}}(\bm O) - \theta_{kt}\varphi_{D_t}(\bm O)\right]$. This decomposition motivates an alternative one-step construction discussed in Section \ref{sec:methods}.

\section{Methods}
\label{sec:methods}

\subsection{Plug-in Estimators}

Given an estimate $\hat g_0(\bX,t)$ of the nuisance function   $g_0(\bX,t)$, one can form plug-in estimators for $\delta_{kt}$ and $\psi_t$ based on the identifying functionals outlined in Theorems \ref{thm:theta-id} and \ref{thm:kl-id}:
% One can construct plug-in estimators $\bm \delta_t$ and $\psi_t$
\begin{align*}
    \hat \psi_t^{\text{PI}} &= \sumn \hat w_0(\bX_i,t) \log \hat w_0(\bX_i,t), \ \ \ \hat w_0(\bX_i,t) = \hat g_0(\bX_i,t)/\hat \mu_{0t}    \\
    \hat{\theta}_{kt}^{\text{PI}} &= \left(\frac{1}{\hat \mu_{0t}} \right) \sumn X_{ki} \cdot \hat g_0(\bX_i,t), \ \ \ \hat \mu_{0t} = \sumn \hat g_0(\bX_i,t),
\end{align*}
where we can collect $\hat{\bm \theta}_t^{\text{PI}} = (\hat{\theta}_{1t}^{\text{PI}}, \ldots, \hat{\theta}_{pt}^{\text{PI}})$. In turn, one can construct plug-in estimators for $\delta_{kt}$:
$
\hat{{\delta}}_{kt}^\text{PI} = (\bar X_k -\hat{\theta}_{kt}^\text{PI})/\widehat{\text{SD}}_k,
$
where $\bar X_k := \sumn X_{ki}$ and $\widehat{\text{SD}}_k \eqdef \sqrt{\frac{1}{n-1}\sum_{i=1}^n (X_{ki} - \bar X_k)^2}$.

In principle, $g_0(\bX,t)$ can be estimated with parametric survival models, or through more flexible nonparametric machine learning methods. While the above plug-in estimators will be consistent provided $\hat g_0(\bX,t)$ is a consistent estimator of $g_0(\bX,t)$, the rate at which they converge will typically be dictated by the behavior of the nuisance function estimator $\hat g_0(\bX,t)$. When $\hat g_0(\bX,t)$ is obtained from flexible machine learning methods to avoid restrictive modeling assumptions, $\hat \delta_{kt}^{\text{PI}}$ and $\hat \psi_t^\text{PI}$ will generally converge at a slower than parametric $\sqrt n$-rate and carry a non-negligible first-order bias---commonly referred to as \textit{plug-in bias}---that complicates the construction of valid confidence intervals \citep{kennedy2024semiparametric}. Although one can specify parametric models to avoid this plug-in bias, parametric plug-in estimators will similarly suffer bias if the  chosen model is misspecified.

\subsection{One-Step Debiased Estimation}

To enable estimation of $\psi_t$ and $\delta_{kt}$ with flexible machine learning methods used for nuisance estimation, we construct one-step debiased estimators by adding the empirical average of each functional's estimated EIF onto their initial plug-in estimator:
\begin{align}
 &   \hat \psi_t^\text{OS} = \hat \psi_t^\text{PI} + \sumn \hat \varphi_{\psi_t}(\bm O_i); \label{eq:psi-onestep} \\
&   \hat{\theta}_{kt}^\text{OS} = \hat {\theta}_{kt}^\text{PI} + \sumn \hat \varphi_{\theta_{kt}}(\bm O_i). \label{eq:theta-onestep-dir}
\end{align}
Given the ratio $\theta_{kt} = N_{kt}/D_t$ discussed in Section \ref{sec:identification}, one can alternatively construct a one-step estimator for $\theta_{kt}$ by constructing one-step estimators for each of $N_{kt}$ and $D_t$ before taking their ratio:
\begin{align}
\hat \theta_{kt}^{\text{OS}} &= \frac{\hat N_{kt}^{\text{OS}}}{\hat D_t^{\text{OS}}}, \label{eq:theta-onestep-ratio} \\
\hat N_{kt}^{\text{OS}} &= \sumn \big[\hat g_0(\bX_i,t) X_{ki} + \tfrac{\II(A_i=0)}{1-\hat e(\bX_i)}(S_{ti} - \hat g_0(\bX_i,t)) X_{ki}\big], \nonumber \\
\hat D_t^{\text{OS}}   &= \sumn \big[\hat g_0(\bX_i,t) + \tfrac{\II(A_i=0)}{1-\hat e(\bX_i)}(S_{ti} - \hat g_0(\bX_i,t))\big]. \nonumber
\end{align}
For either choice of $\hat \theta_{kt}^\text{OS}$, the corresponding one-step estimator of $\delta_{kt}$ is then $\hat \delta_{kt}^{\text{OS}} = (\bar X_k - \hat \theta_{kt}^{\text{OS}})/\widehat{\text{SD}}_k.$ Although both versions of $\hat \theta_{kt}^\text{OS}$ above will be asymptotically equivalent under consistent and sufficiently fast estimation of $g_0(\bX,t)$ and $e(\bX)$, the two estimators can exhibit meaningfully different behavior under misspecified working models for these two nuisance functions. Throughout this section and Sections \ref{sec:simulation} and \ref{sec:application}, we primarily consider the ratio formulation in \eqref{eq:theta-onestep-ratio}, which endows $\hat \theta_{kt}^\text{OS}$ with an additional robustness against potential misspecification of $g_0(\bX,t)$ relative to the one-step estimator in Equation \eqref{eq:theta-onestep-dir}. A similar robustness distinction between direct and ratio one-step estimators has been noted for the causal risk ratio \citep{boughdiri2025quantifying}.

An important property of one-step estimators, and debiased machine learning estimators more broadly, is that they possess tractable asymptotic distributions under often mild conditions on the estimation rates of their respective nuisance functions. This allows flexible machine learning methods to be used for nuisance estimation while still permitting valid inference. We note that one-step estimators are often referred to as ``doubly robust," in part because the well-known augmented inverse probability weighting (AIPW) estimator for cross-sectional average treatment effects is a one-step estimator that possesses a model double-robustness property: it remains consistent if either the outcome regression or the propensity score model is correctly specified \citep{kurz2022augmented}. In general, however, the conditions required for consistency and asymptotic normality of one-step estimators will be tied to the specific functional being estimated. Let $|| \hat f - f||_2 = \left[\E\left\{\bigl(\hat f(\bm O)-f(\bm O)\bigr)^2\right\}\right]^{1/2}$ denote the $L_2$ estimation error of a generic nuisance function estimator $\hat f$. In turn, we characterize the  asymptotic distributions of our one-step estimators below, first focusing on $\hat \theta_{kt}^\text{OS}$:
\begin{theorem}
\label{thm:theta-asym}
 Suppose Assumptions \ref{as:consistency}-\ref{as:monotonicity} hold, that $\hat g_0$ and $\hat e$ are obtained via cross-fitting, and that $1 - \hat e(\bX)$ is bounded away from zero almost surely. Further assume that $\mu_{0t}>0$ and that each $X_k$ is bounded almost surely. Then,
\begin{enumerate}
\item[(i)] \emph{(Consistency.)} If $\|\hat g_0 - g_0\|_2 = o_\PP(1)$ or $\|\hat e - e\|_2 = o_\PP(1)$, then $\hat \theta_{kt}^{\text{OS}} \xrightarrow{p} \theta_{kt}$.
\item[(ii)] \emph{(Asymptotic linearity.)} If both $\|\hat g_0 - g_0\|_2 = o_\PP(1)$ and $\|\hat e - e\|_2 = o_\PP(1)$, then
\[
\hat \theta_{kt}^\text{OS} - \theta_{kt} = \sumn \varphi_{\theta_{kt}}(\bm O_i) + o_\PP(n^{-1/2}) + O_\PP(R_{n,\theta_{kt}}), \qquad R_{n,\theta_{kt}} = \|\hat g_0 - g_0\|_2 \cdot \|\hat e - e\|_2.
\]
In particular, if $R_{n,\theta_{kt}} = o_\PP(n^{-1/2})$ then $\sqrt{n}(\hat \theta_{kt}^{\text{OS}} - \theta_{kt}) \xrightarrow{d} \mathcal{N}(0, \E[\varphi_{\theta_{kt}}(\bm O)^2])$.
\end{enumerate}
\end{theorem}

Theorem \ref{thm:theta-asym} establishes that $\hat \theta_{kt}^{\text{OS}}$ is doubly robust in a traditional sense. Specifically, $\hat \theta_{kt}^{\text{OS}}$ is consistent provided one of the two estimated nuisance functions $\hat g_0(\bX,t)$ and $\hat e(\bX)$ is consistent, and  $\sqrt n$ consistent and asymptotically normal provided the product of nuisance estimator convergence rates of $\hat g_0(\bX,t)$ and $\hat e(\bX)$ is $o_\PP(n^{-1/2})$, which holds, for example, when both nuisance estimators converge at $o_P(n^{-1/4})$ rates---a condition satisfied by many modern flexible regression methods \citep{kennedy2024semiparametric}. 
Given that $\bar X_k$ and $\widehat{\text{SD}}_k$ will tend to be $\sqrt n$ consistent and asymptotically normal under mild regularity conditions, the conditions for asymptotic normality of $\hat \delta_{kt}^\text{OS}$ are analogous to those for $\hat \theta_{kt}^\text{OS}$ and are an immediate corollary of Theorem \ref{thm:theta-asym}. Specifically, under the same conditions listed in Theorem \ref{thm:theta-asym} we have that
\[
\hat \delta_{kt}^\text{OS} - \delta_{kt} = \sumn \varphi_{\delta_{kt}}(\bm O_i) + o_\PP(1/\sqrt n) + O_\PP(||\hat g_0 - g_0 ||_2 \cdot ||\hat e - e||_2), \ \ \ \text{where} 
\]
\[
\varphi_{\delta_{kt}}(\bm O) = \frac{1}{\text{SD}_k}\left[ (X_k - \E[X_k]) - \varphi_{\theta_{kt}}(\bm O) - \delta_{kt} \frac{(X_k - \E[X_k])^2 - \text{SD}_k^2}{2\text{SD}_k} \right]
\]
is the influence function for $\delta_{kt}$. In the coming Section we discuss how these conditions imply a straightforward means to construct confidence intervals for $\delta_{kt}$, after discussing similar conditions that allow us to characterize the asymptotic distribution of $\hat \psi_t^\text{OS}$:
\begin{theorem}
\label{thm:psi-asym}
Suppose Assumptions \ref{as:consistency}--\ref{as:monotonicity} hold, that $\hat g_0(\bX,t)$ and $\hat e(\bX)$ are obtained via cross-fitting, and that $1 - \hat e(\bX)$ is bounded away from zero almost surely. Additionally assume $\mu_{0t}, w_0(\bX,t)$ and $\hat w_0(\bX,t)$ are all bounded away from 0 almost surely. Then,
\begin{enumerate}
\item[(i)] \emph{(Consistency.)} If $\|\hat g_0 - g_0\|_2 = o_\PP(1)$, then $\hat \psi_t^{\text{OS}} \xrightarrow{p} \psi_t$, regardless of whether $\hat e$ is consistent.
\item[(ii)] \emph{(Asymptotic linearity.)} If, in addition, $\|\hat e - e\|_2 = o_\PP(1)$, then
\[
\hat \psi_t^{\text{OS}} - \psi_t = \sumn \varphi_{\psi_t}(\bm O_i) + o_\PP(n^{-1/2}) + O_\PP(R_{n,\psi_t}), \ \ \  R_{n,\psi_t} = \|\hat g_0 - g_0\|_2 \cdot \|\hat e - e\|_2 + \|\hat g_0 - g_0\|_2^2.
\]
In particular, if $R_{n,\psi_t} = o_\PP(n^{-1/2})$, then $\sqrt{n}(\hat \psi_t^{\text{OS}} - \psi_t) \xrightarrow{d} \mathcal{N}(0, \E[\varphi_{\psi_t}(\bm O)^2])$.
\end{enumerate}
\end{theorem}

Relative to the second-order remainder $R_{n,\theta_{kt}}$ for $\hat \theta_{kt}^\text{OS}$, the second-order remainder $R_{n,\psi_t}$  carries the additional term $\|\hat g_0 - g_0\|_2^2$, which arises due to the nonlinear transformation of the nuisance $w_0(\bX,t)$ in \eqref{eq:psi-id}. As a consequence, $\hat \psi_t^{\text{OS}}$ requires the stronger condition $\|\hat g_0 - g_0\|_2 = o_\PP(n^{-1/4})$ to attain $\sqrt n$-consistency, regardless of how quickly $\hat e(\bX)$ converges. 
%This is a common drawback of nonlinear functionals. 
A practical implication is that even in randomized trials, where $e(\bX)$ is known, $\hat \psi_t^{\text{OS}}$ continues to require $n^{-1/4}$-consistent estimation of $g_0$, while $\hat \theta_{kt}^{\text{OS}}$ and $\hat \delta_{kt}^{\text{OS}}$ enjoy a strictly stronger robustness guarantee, which we discuss in Section \ref{sec:rand-trial}. 

We note that $\psi_t$ is one instance of a broader class of scalar divergence measures of the form $D_{h,t} = \E[h(w_0(\bX,t))]$, indexed by a smooth transformation $h$ of $w_0(\bX,t)$. Notably, the KL divergence corresponds to $h(w) = w \log w$, whereas other common divergence measures like the $\chi^2$ and squared Hellinger distances correspond to $h(w) = (w-1)^2$ and $h(w) = \tfrac12(\sqrt w - 1)^2$, respectively. The same plug-in and one-step estimation framework  used for $\hat \psi_t^\text{OS}$ applies to these broader functionals, and in the Supplementary Material we give the general form of their efficient influence function. Importantly, the stronger rate requirement for $\hat \psi_t^\text{OS}$ similarly holds for the distributional divergence measures in this broader class, implying this weaker robustness property is an inherent cost in estimating functionals that capture higher-order distributional differences. 

\subsection{Inference}

Recall that when the conditions of Theorems \ref{thm:theta-asym} and \ref{thm:psi-asym} are satisfied, $\hat \delta_{kt}^\text{OS}$ and $\hat \psi_t^\text{OS}$ are consistent and asymptotically normal with variances equal to the variance of each estimator's respective EIF. Critically, this implies that one can construct asymptotic Wald-type confidence intervals for each of $\theta_{kt}$, $\delta_{kt}$, and $\psi_t$ from the empirical variance of each functional's corresponding estimated influence functions. For a generic estimand $\eta \in \{\theta_{kt}, \delta_{kt}, \psi_t\}$ with corresponding one-step estimator $\hat \eta^{\text{OS}}$, one can construct a $(1-\alpha)$ confidence interval through
\[
\hat \eta^{\text{OS}} \pm z_{1-\alpha/2} \cdot \hat \sigma_\eta / \sqrt n, \qquad \hat \sigma_\eta^2 = \sumn (\hat \varphi_\eta(\bm O_i) - \bar{\hat \varphi}_\eta)^2,
\]
where $\bar{\hat \varphi}_\eta = \sumn \hat \varphi_\eta(\bm O_i)$. One can alternatively obtain confidence intervals through bootstrapping, which can exhibit stronger coverage in small samples. Particularly when the number of covariates $p$ or time points $T$ are large, one can adjust $\alpha$ to control family-wise error rates \citep{dmitrienko2013traditional}, though when interest lies in omnibus testing for general distributional differences, we recommend a nonparametric test of $S_t\indep\bX \mid A=0$, since this independence implies $\psi_t=0$. In particular, the Wald intervals for $\psi_t$ are intended to estimate its magnitude rather than test $H_0: \psi_t=0$, since the usual $\sqrt n$ asymptotic distribution of the one-step estimator is degenerate under the null.

\subsection{Special Case: Randomized Trial}
\label{sec:rand-trial}
The conditions in Theorems \ref{thm:theta-asym} and \ref{thm:psi-asym} accommodate both non-randomized and randomized studies, where the treatment propensity scores $e(\bX)$ are known by design. Theorem \ref{thm:theta-asym} suggests that when $e(\bX)$ is known, one can construct consistent and asymptotically normal estimators of $\theta_{kt}$ regardless of the accuracy in estimating $g_0(\bX,t)$. Randomized studies typically have small to moderate sample sizes, implying one may prefer to specify parsimonious, but likely misspecified, parametric working models  for $g_0(\bX,t)$.
%governed by a finite-dimensional parameter $\bm \beta$.
Analogous to results ensuring the validity of the analysis-of-covariance (ANCOVA) estimator for estimating treatment effects in randomized studies \citep{tsiatis2008covariate}, the following Proposition establishes that one can conduct valid inference on $\theta_{kt}$ regardless of the model specified for $g_0(\bX,t)$.
\begin{proposition}
    \label{prop:work-mod-asym}
    Suppose the conditions of Theorem \ref{thm:theta-asym} hold, and additionally that $e(\bX)$ is known by design. Further suppose that $\hat g_0(\bX,t) \rightarrow g^*_0(\bX,t)$, where $g^*_0(\bX,t)$ may not necessarily equal $g_0(\bX,t)$.   Then $\hat \theta_{kt}^{\text{OS}}$ is consistent and asymptotically linear for $\theta_{kt}$ regardless of whether $\hat g_0$ is consistent for $g_0$, with
    \[
    \hat \theta_{kt}^{\text{OS}} - \theta_{kt} = \sumn \varphi_{\theta_{kt}}^*(\bm O_i) + o_\PP(n^{-1/2}),
    \]
    where $\varphi_{\theta_{kt}}^*$ denotes the influence function evaluated at  $g^*(\bX,t)$.
\end{proposition}
Importantly, Proposition \ref{prop:work-mod-asym} implies that when $e(\bX)$ is known, there are infinitely many asymptotically linear estimators for $\theta_{kt}$, all indexed by the limit $g^*(\bX,t)$  of the nuisance estimator $\hat g_0(\bX,t)$. Along with implying any choice of $\hat g_0(\bX,t)$ will yield a consistent estimator of $\theta_{kt}$, this critically implies one can continue to construct valid Wald-type confidence intervals for $\theta_{kt}$ and $\delta_{kt}$ based on the empirical variance of $\hat \varphi_{\theta_{kt}}^*(\bm O)$, evaluated at any working model for $g_0$. Unfortunately, the analogous robustness result does not hold for $\hat \psi_t^{\text{OS}}$. Recalling the form of $R_{n,\psi_t}$ from Theorem \ref{thm:psi-asym}, even when $e$ is known the second-order term $\|\hat g_0 - g_0\|_2^2$ in $R_{n,\psi_t}$ implies that  $\sqrt n$-consistency of $\hat \psi_t^{\text{OS}}$ continues to require $\|\hat g_0 - g_0\|_2 = o_\PP(n^{-1/4})$. In randomized trial settings,  SMD estimation therefore offers a meaningfully stronger robustness guarantee than the KL divergence and more general functionals involving smooth transformations of $w_0(\bX,t)$.

In principle, one can estimate $g_0(\bX,t)$ separately at each $t$, or pooled across $t$. The EIF for $\theta_{kt}$ is derived under a model that imposes no restrictions on the joint distribution of $(S_t,Y_t)$ across time, but pooling may nevertheless improve finite-sample precision by borrowing information across time points.  Specifically Proposition \ref{prop:work-mod-asym} implies the asymptotic variance of $\hat \theta_{kt}^\text{OS}$ is determined by the probability limit of the nuisance estimator $\hat g_0(\bX,t)$, meaning if separate pooled and time-indexed estimators of $g_0(\bX,t)$ are both correctly specified, the resulting $\hat \theta_{kt}^\text{OS}$ will have the same asymptotic variance under either nuisance estimator. Relative to a fully time-indexed approach, any finite-sample gains from pooling depend on adequately specifying $g_0(\bX,t)$'s dependence on $t$, meaning misspecification could result in losses to efficiency.

\section{Relaxing Monotonicity}
\label{sec:relaxing-monotonicity}
Although the monotonicity Assumption \ref{as:monotonicity} will be appropriate in many biomedical settings, its validity will tend to vary across applications, and for many disease areas may be a particularly strong condition. Particularly in ALS research, where previous therapies have demonstrated possible harmful effects on disease progression \citep{meininger2006pentoxifylline,gordon2007efficacy}, there may be a desire to reduce reliance on the monotonicity assumption when assessing newer investigational treatments.

To address this limitation, in this Section we consider sensitivity analyses that allow one to partially relax Assumption \ref{as:monotonicity}. To this end, we consider the time-indexed sensitivity function
\[
\rho(\bX,t) \eqdef \PP(S_t(1)=0 \mid S_t(0)=1,\bX),
\]
the $\bX$-conditional probability of treatment causing death by time $t$. Monotonicity is satisfied when $\rho(\bX,t)=0$ almost surely, and large values of $\rho(\bm x,t)$ imply severe violations of monotonicity within covariate strata $\bm x$. For brevity, we discuss how $\rho(\bX,t)$ can be used to enable partial identification of $\theta_{kt}$, providing an analogous strategy for $\psi_t$ in the Supplementary Material.
In settings where $\rho(\bX,t)$ is constant in $\bX$, the following Proposition establishes that the previous identifying functional \eqref{eq:theta-id} for $\theta_{kt}$ remains valid.
\begin{proposition}
\label{prop:const-mono}
Suppose $\rho(\bX,t) = p_t \in [0,1)$ almost surely. Then under Assumptions \ref{as:consistency}-\ref{as:surv-unc}, $\theta_{kt}$ continues to be identified by
\[
\theta_{kt} = \theta_{kt}^\text{mono} := \frac{\E[X_k \cdot  g_0(\bX,t)]}{\E[g_0(\bX,t)]} = \E[X_k \cdot w_0(\bX,t)].
\]
\end{proposition}
Proposition \ref{prop:const-mono} implies that violations to Assumption \ref{as:monotonicity} are permissible, in the sense that identification of $\theta_{kt}$ and $\psi_t$ remains valid at any $t$ so long as $\rho(\bX,t)$ is constant over $\bX$. An analogous result has been demonstrated for identification of $\tau_t$ \citep{zehavi2023matching}. Importantly, this implies our proposed methods will incur bias if $\rho(\bX,t)$ varies strongly in $\bX$, suggesting sensitivity analyses can be informed by specifying the variability of $\rho(\bX,t)$.
Continuing to let $\theta_{kt}^\text{mono} = \E[X_k \cdot w_0(\bX,t)]$, in the Supplementary Material we establish the following result:
\begin{proposition}
\label{prop:sens-bias}
Let $\sigma_{\rho_t} := \text{SD}(\rho(\bX,t) \mid S_t{(0)=1})$ and $\bar \rho_t := \PP(S_t(1)=0|S_t(0)=1)$, and suppose $\bar \rho_t < 1$. Then, under Assumptions \ref{as:consistency}-\ref{as:surv-unc},
\begin{equation}
\label{eq:sens-bound}
 |\theta_{kt} - \theta_{kt}^\text{mono}| \leq \frac{\sigma_{\rho_t} \cdot \text{SD}(X_k \mid  S_t(0)=1)}{1-\bar \rho_t}.
\end{equation}
\end{proposition}

Proposition \ref{prop:sens-bias} suggests a more practical means to investigate violations to Assumption \ref{as:monotonicity}. Rather than specify the entire sensitivity function $\rho(\bX,t)$, a typically intractable task, \eqref{eq:sens-bound} suggests one can instead specify a grid of plausible values for the scalar parameters  $\bar \rho_t$ and $\sigma_{\rho_t}$, which capture the overall rate of monotonicity violations and the degree to which they vary among units with $S_t(0)=1$, respectively.  For fixed values of $\sigma_{\rho_t}$ and $\bar \rho_t$, noting that \eqref{eq:sens-bound} implies
\[
\theta_{kt} \in \left[\theta_{kt}^\text{mono} - \frac{\sigma_{\rho_t} \cdot \text{SD}(X_k \mid S_t(0)=1)}{1-\bar \rho_t},\ \  \theta_{kt}^\text{mono} +\frac{\sigma_{\rho_t} \cdot \text{SD}(X_k \mid S_t(0)=1)}{1-\bar \rho_t}
\right],
\]
one can in turn construct debiased one-step estimators of the bounds $\theta_{kt}^\text{L}, \ \theta_{kt}^U$ of $\theta_{kt}$:
\begin{align}
( \hat \theta_{kt}^\text{L-OS},\ \hat \theta_{kt}^\text{U-OS} ) = 
\hat \theta_{kt}^\text{OS} \pm \frac{\sigma_{\rho_t} \cdot \hat s_{kt}^\text{OS}}{1-\bar \rho_t},
\label{eq:sens-bounds-est}
\end{align}
where $\hat s_{kt}^\text{OS}$ is a one-step estimator for $s_{kt} := \text{SD}(X_k \mid S_t(0)=1)$, which can be identified by a functional of the observed data using similar arguments to those in Section \ref{sec:identification}. Full details on the estimation procedure for obtaining $\hat s_{kt}^\text{OS}$, and in turn the bounds $\hat \theta_{kt}^\text{L-OS}$ and $\hat \theta_{kt}^\text{U-OS}$, are provided in the Supplementary Material.
The above bounds depend on both $\bar\rho_t$ and $\sigma_{\rho_t}$, where both scalar parameters can be calibrated from subject-matter knowledge or historical data. The two parameters are restricted by $\sigma_{\rho_t} \leq \sqrt{\bar\rho_t(1-\bar\rho_t)}$, since $\rho(\bX,t) \in [0,1]$. In practice, one can calibrate a plausible upper bound on $\bar\rho_t$ and report a sensitivity curve plotting $\hat \theta_{kt}^{\text{OS}}$ together with the DR bounds across a grid of $\sigma_\rho$ values, with the shape of the curve communicating how robust the substantive conclusion---e.g., that $|\delta_{kt}| > 0.1$ at a given $t$---is to plausible covariate-dependent variation in $\rho(\bX,t)$.  We demonstrate this approach in Sections \ref{sec:simulation} and \ref{sec:application}.

\section{Simulation}
\label{sec:simulation}

\subsection{Setup}
In this Section, we assess the finite-sample performance of our proposed methods in a range of settings designed to mimic features of longitudinal randomized trials. Baseline covariates and treatment are drawn as $(X_1,X_2,X_3)\overset{\text{iid}}{\sim}N(0,1)$ and $\PP(A=1\mid\bX)=1/2$, where $\bX=(X_1,X_2,X_3)$. For each subject, potential survival outcomes are generated sequentially through a discrete-time hazard function
\[
\lambda_a(\bX,t) = \PP(S_t(a)=0 \mid S_{t-1}(a)=1,\bX),
\]
so any subject who would survive through time $t-1$ under treatment $A=a$  has a $1-\lambda_a(\bX,t)$ probability of surviving through time $t$ under treatment $A=a$. In turn, notice $\PP(S_t(a)=1\mid \bX) = \prod_{s=1}^t \{1-\lambda_a(\bX,s)\}$. For each subject and time point $t$, if $S_{t-1}(a)=1$ or $t=1$ we draw $U_t\sim\text{Uniform}(0,1)$, setting $S_t(a)=\mathbb{I}\{U_t \geq \lambda_a(\bX,t)\}$. For $t>1$, if $S_{t-1}(a)=0$  we set $S_t(a)=0$. To ensure monotonicity, we set $\lambda_1(\bX,t)=0.6\cdot \lambda_0(\bX,t)$, noting in turn $U_t \geq \lambda_0(\bX,t) \implies U_t \geq \lambda_1(\bX,t)$. Since $\lambda_0(\bX,t) = 1-g_0(\bX,t)/g_0(\bX,t-1)$, we parameterize $g_0(\bX,t)$ directly by setting 
\[
g_0(\bX,t)=\PP\big(S_t(0)=1\mid\bX\big)=\text{expit}\left(\beta_0-\beta_1\frac{t}{T}-\sum_{k=1}^3\alpha_k X_k-\sum_{k=1}^3\kappa_k X_k^2-\sum_{j<k}\gamma_{jk}X_j X_k\right),
\]
for $t>0$, letting $g_0(\bX,0)=1$ and implying $\lambda_a(\bX,t)$ is indirectly parameterized through $g_0(\bX,t)$. 

Additionally, we generate survival-conditional potential outcomes according to
\[
Y_t(0) = 3 - \bar X/10 - \delta^t\bar X/5 - \delta^t \bar Q/15, \ \ \ Y_t(1) = 4 - \bar X/10-\delta^t\bar X/8-\delta^t\bar Q/15.
\]
where $\bar X = (X_1+X_2+X_3)/\sqrt 3$, $\bar Q = \{(X_1^2-1) + (X_2^2-1)\}/\sqrt 2$ and $\delta=1.05$. Since our primary focus is on estimating $\bm \delta_t$ and $\psi_t$, $Y_t(1)$ and $Y_t(0)$ are only used to illustrate how survival measures can accompany $\tau_t$.

\begin{figure}[t!]
\centering
\includegraphics[width=0.65\textwidth]{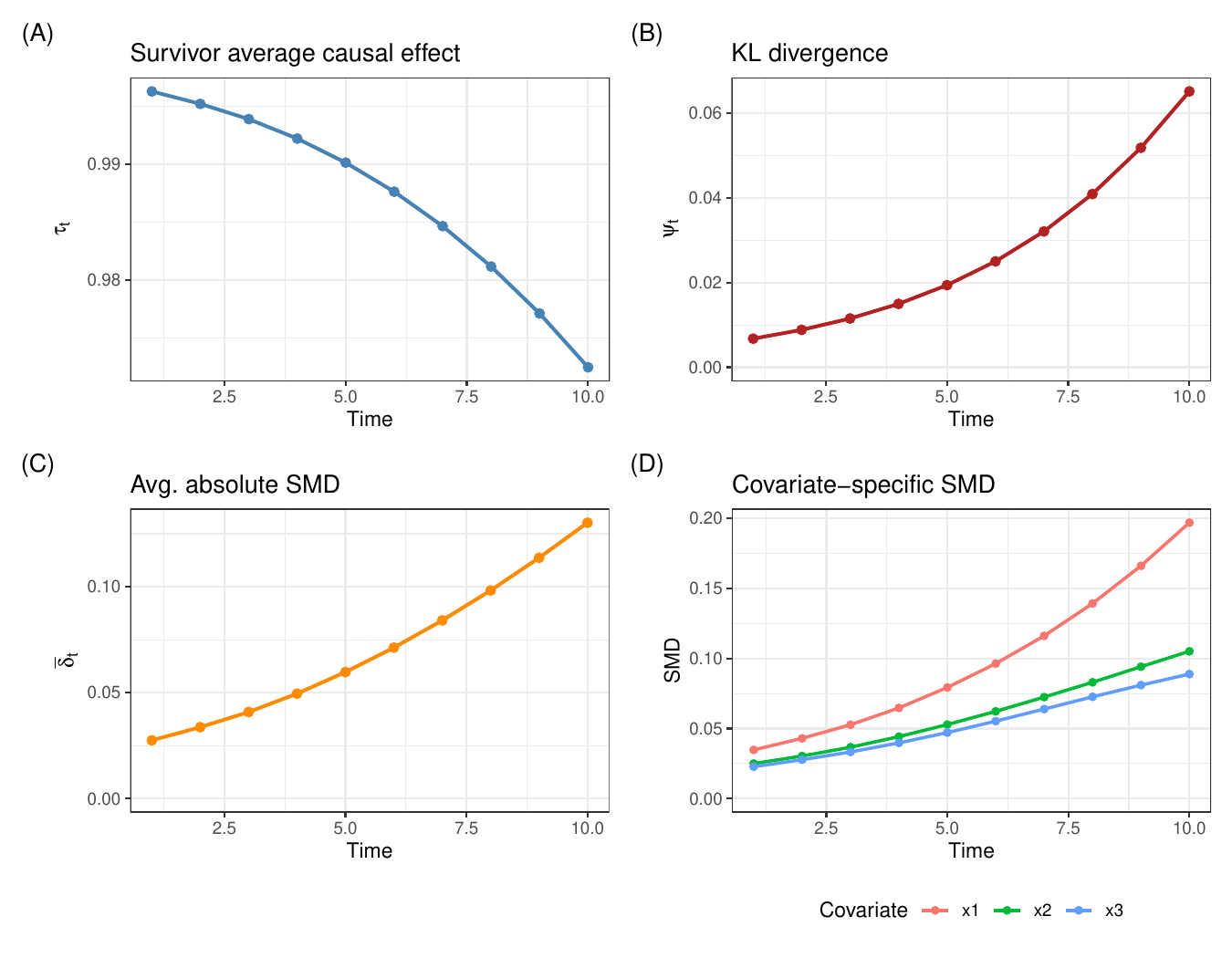}
\caption{Trajectories of causal estimands and divergence measures induced by the simulation data-generating process, as functions of follow-up time $t$. (A) survivor average causal effect $\tau_t$; (B) KL divergence $\psi_t$; (C) average absolute SMD $\bar{\bm \delta}_t$; (D) per-covariate SMDs $\delta_{kt}$.
\ifbiostatistics
\\ \textbf{Alt text}: Four-panel plot showing true simulation estimand values over time. Plot shows a small decline in the true SACE over time, with both the KL divergence and standardized mean differences increasing over time.
\fi}
\label{fig:oracle}
\end{figure}

Figure \ref{fig:oracle} displays the true  $\tau_t,\bm \delta_t$, and $\psi_t$ induced by this data-generating process. Both $\delta_{kt}$ and $\psi_t$ increase in $t$, implying  the time-$t$ survivor principal stratum becomes increasingly less representative of the baseline covariate distribution as the trial progresses.  For reference, $|\delta_{kt}|=0.1$ and $|\delta_{kt}|=0.2$ are often used as thresholds for moderate divergence in the matching and weighting literatures \citep{austin2009balance}, while KL divergences ranging from $10^{-3},10^{-2}$, and $10^{-1}$ have been informally described as small, moderate and large divergences, respectively \citep{commenges2008estimating}. Relatedly, $\tau_t$ varies over $t$ in this setting, while the composition of the survivor stratum also changes over follow-up. An analyst inspecting only $\tau_t$ in this setting would observe modest changes in the estimated treatment effect over follow-up, with no indication that the underlying survivor stratum is drifting substantially from the baseline population.

To investigate the robustness to model misspecification of our proposed one-step estimators in finite samples, we report results from estimators with correct and incorrect specifications for $g_0(\bX,t)$. At each time point $t$,  in the \textit{correct specification} scenario we estimate $g_0(\bX,t)$ using a SuperLearner \citep{van2007super} ensemble of (i) logistic regression models with pairwise interactions and main effects terms, (ii) ridge-penalized logistic regression with main effects, interaction, and quadratic terms, and (iii) second-degree multivariate adaptive regression splines \citep{milborrow2018earth} (MARS). For the \textit{incorrect specification} scenario, we estimate $g_0(\bX,t)$ with a main effects logistic regression, importantly omitting covariate interactions and higher-order main effects. We use five-fold cross-fitting \citep{kennedy2024semiparametric} to construct $\hat \psi_t^\text{OS}$ and $\hat \delta_{kt}^\text{OS}$ with the above learners. Full details on our estimation procedure are provided in the Supplementary Material.

\subsection{Results}

\begin{figure}[t!]
\centering
\includegraphics[width=0.95\textwidth]{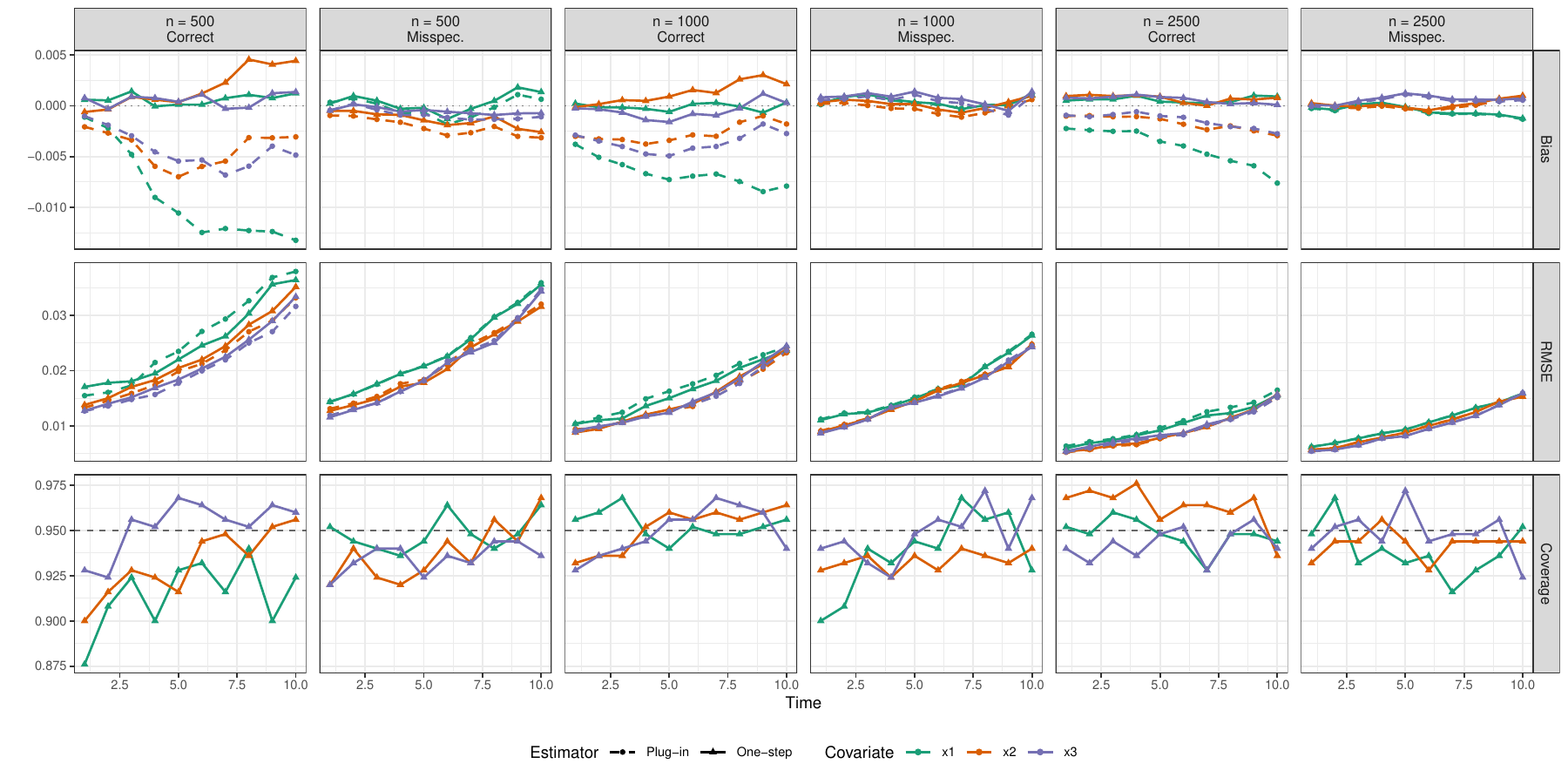}
\caption{Estimation performance for covariate SMDs $\delta_{kt}$ under correct and incorrect specification of $g_0(\bX,t)$,  varying $n \in \{500, 1000, 2500\}$.
\ifbiostatistics
\\ \textbf{Alt text}: Bias, root mean squared error, and coverage for SMD estimators, varying the sample size and whether the survival-model is correctly specified. The one-step estimators remain approximately unbiased with approximately nominal coverage regardless of specification.
\fi
}
\label{fig:smd-perf}
\end{figure}

\begin{figure}[t!]
\centering
\includegraphics[width=1\textwidth]{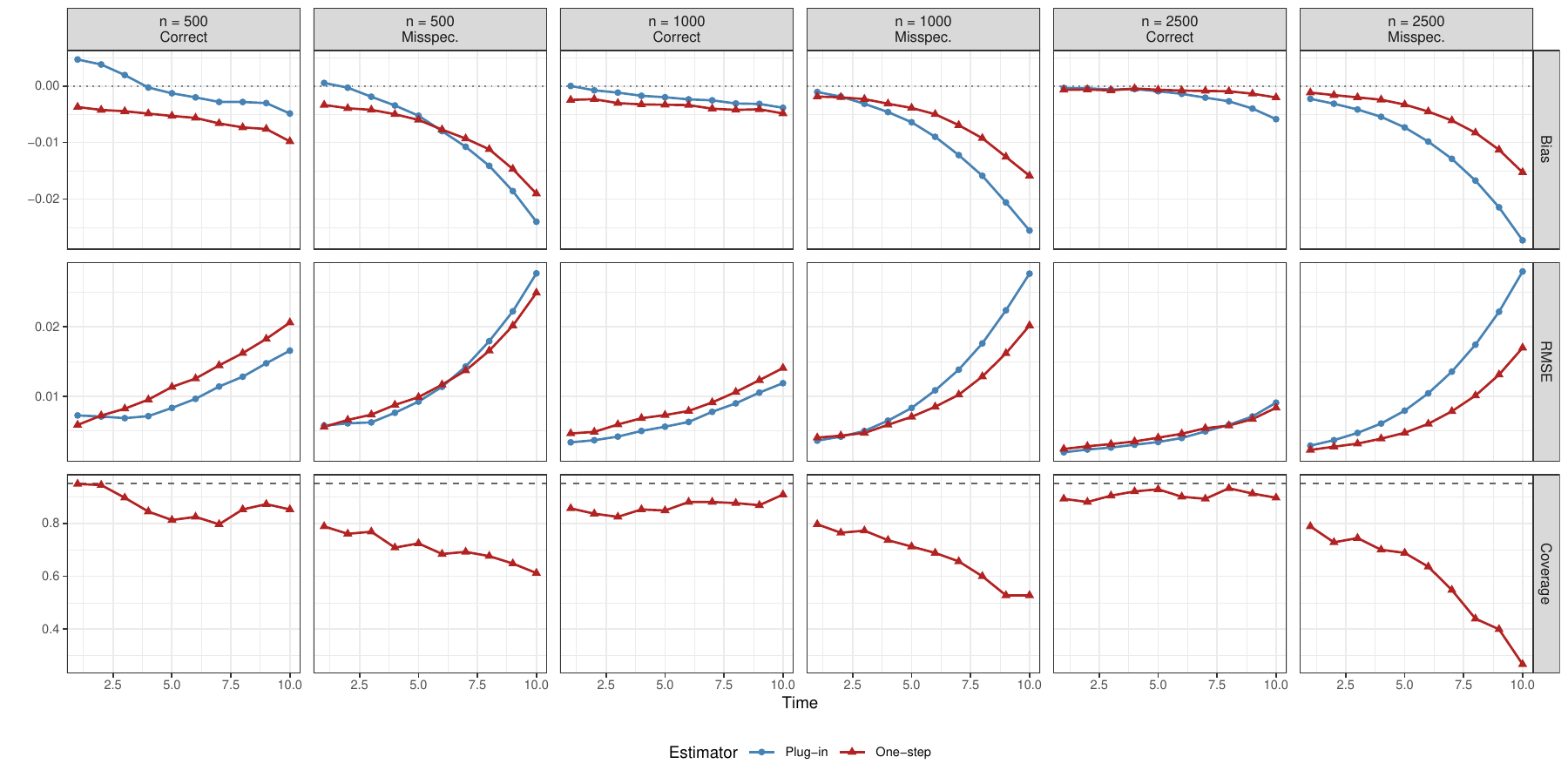}
\caption{Estimation performance for the KL divergence $\psi_t$ under correct and incorrect specification of $g_0(\bX,t)$, varying $n \in \{500, 1000, 2500\}$.
\ifbiostatistics
\\ \textbf{Alt text}: Bias, root mean squared error, and coverage for SMD estimators, varying the sample size and whether the survival-model is correctly specified. The one-step estimators are slightly biased and miscover under misspecification, though are approximately unbiased with near-nominal coverage at larger sample sizes under correct specification. 
\fi
}
\label{fig:kl-perf}
\end{figure}

Figures \ref{fig:smd-perf} and \ref{fig:kl-perf} display the performance of our proposed divergence estimators. We additionally assess coverage of the proposed sensitivity-bound estimators under monotonicity violations below in Figure \ref{fig:sens-sim}.

\paragraph{Divergence measures.} We first focus on Figures \ref{fig:smd-perf} and \ref{fig:kl-perf}. Notably, $\hat \delta_{kt}^\text{OS}$ is approximately unbiased for each covariate $k$, regardless of model specification, while $\hat \delta_{kt}^\text{PI}$ exhibits plug-in bias when $g_0(\bX,t)$ is fit with the flexible ensemble learner. Under the misspecified nuisance model, $\hat\delta_{kt}^{\mathrm{PI}}$ is also approximately unbiased in this particular data-generating process, though unlike $\hat\delta_{kt}^{\mathrm{OS}}$, this behavior is not guaranteed by the randomized-trial robustness result.  Importantly, $\hat \delta_{kt}^\text{OS}$ attains approximately valid coverage across all sample sizes, regardless of whether $g_0(\bX,t)$ is correctly specified, consistent with the discussion in Section \ref{sec:rand-trial}.
In contrast, both $\hat \psi_t^\text{PI}$ and $\hat \psi_t^\text{OS}$ exhibit instability and finite-sample bias at smaller sample sizes, particularly at longer time horizons $t$. For larger sample sizes, the one-step estimator demonstrates a degree of protection from misspecification of $g_0(\bX,t)$ relative to $\hat \psi_t^\text{PI}$, though its 95\% confidence intervals exhibit severe miscoverage under misspecification. This is consistent with Theorem \ref{thm:psi-asym}, demonstrating that the KL divergence does not share the same robustness to misspecification of $g_0(\bX,t)$ as the SMDs.

\paragraph{Sensitivity analyses} To assess the sensitivity bound estimators developed in Section \ref{sec:relaxing-monotonicity}, we modify the above data-generating process so that monotonicity is violated. At each time $t$, we set $\rho(\bX,t) = 0.30 \cdot (t/T) \cdot \expit(0.5 X_1)$ so that $S_t(1) = 0$ when $S_t(0) = 1$ for a subset of units across all $t$. Since $\rho(\bX,t)$ varies in $X_1$, notably $\theta_{kt}\neq \theta_{kt}^\text{mono}$. The severity of the monotonicity violation grows with $t$, so that at $t=10$, the true parameters are $\bar \rho_t \approx 0.14$ and $\sigma_\rho \approx 0.03$. Recalling that in practice these sensitivity parameters are unknown and set by the analyst, we fix an  upper bound for the sensitivity parameter $\bar \rho_t = 0.15$ and construct one-step estimators of the sensitivity bounds implied by \eqref{eq:sens-bounds-est} for $\sigma_\rho \in \{0, 0.025, 0.05, 0.075, 0.10, 0.15, 0.20\}$.

\begin{figure}[t!]
\centering
\includegraphics[width=0.95\textwidth]{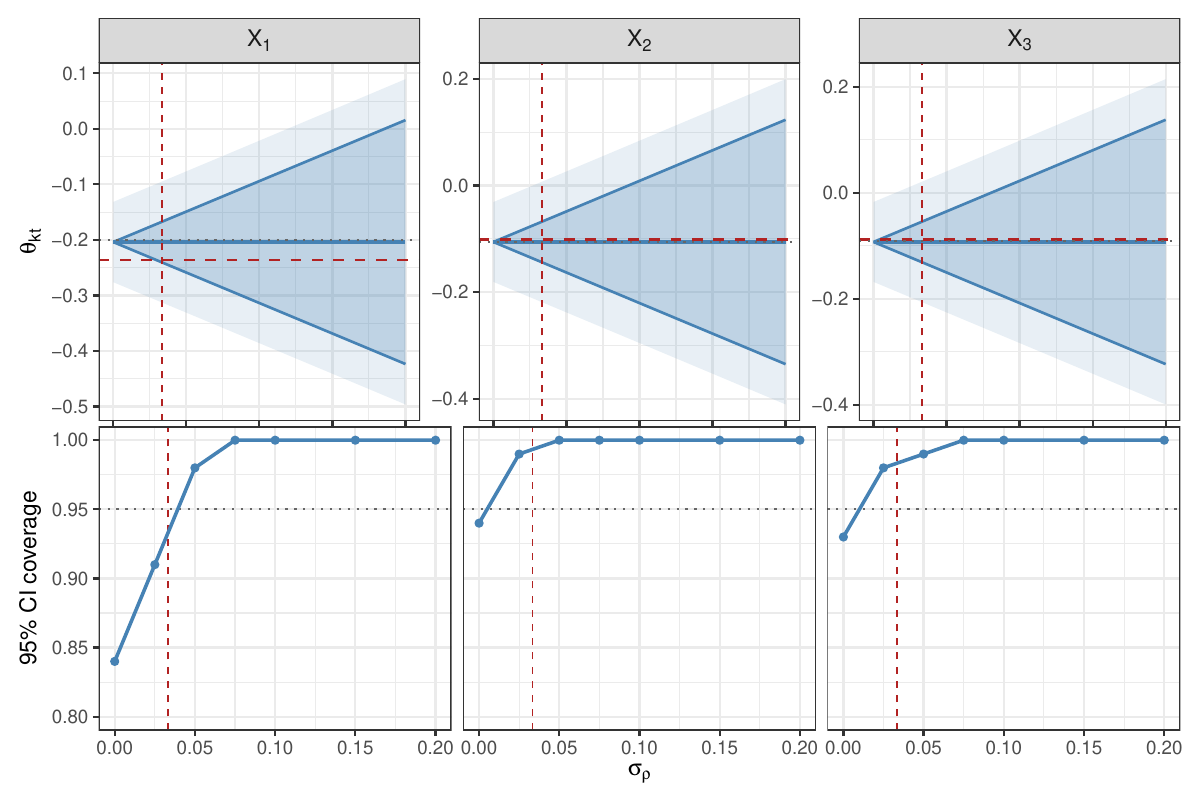}
\caption{\textbf{Top panel}: Sensitivity analysis bounds for $\theta_{kt}$ at $t=10$ with $n=1000$ and $\bar \rho_t = 0.15$. Blue line: mean one-step point estimate $\hat \theta_{kt}^{\text{OS}}$. Inner dark blue ribbon: mean lower and upper sensitivity bounds at the indicated $\sigma_\rho$. Outer light blue ribbon: mean estimated 95\% confidence interval, accounting for estimation uncertainty for point bounds. Red dashed horizontal line: true $\theta_{kt}$. Gray dotted horizontal line: monotonicity-implied $\theta_{kt}^\text{mono}$. Red dashed vertical line: true $\sigma_\rho \approx 0.03$. \textbf{Bottom panel}: 95\% CI coverage based on estimated point bounds.
\ifbiostatistics
\\ \textbf{Alt text:}  Estimated sensitivity bounds for three survivor covariate means. Bounds widen with as the variance sensitivity parameter increases.
\fi
}
\label{fig:sens-sim}
\end{figure}

Figure \ref{fig:sens-sim} displays the results of the sensitivity analysis. The average point bounds widen with the sensitivity parameter $\sigma_\rho$, and for $X_1$---which is the main covariate impacted by monotonicity violations---the average bounds begin to cover $\theta_{kt}$ at the true value $\sigma_\rho \approx 0.03$. The average point bounds for $X_2$ and $X_3$, which are largely unaffected by the monotonicity violation given the form of $\rho(\bX,t)$, similarly grow with $\sigma_\rho$ and always contain their respective target parameter. Continuing to focus on $X_1$, the sensitivity bound 95\% CIs undercover at smaller sensitivity values, while attaining approximately nominal coverage at the true $\sigma_\rho$, and providing conservative overcoverage for larger sensitivity values. Relatedly for $X_2$ and $X_3$, coverage is conservative for all non-zero values of $\sigma_\rho$ considered.

\section{Application: PRO-ACT}
\label{sec:application}

We apply our proposed methods to data from the PRO-ACT ALS database \citep{atassi2014pro} introduced in Section \ref{sec:introduction}. PRO-ACT is a fully anonymized open-access database pooling placebo and treatment-arm data from 13,717 longitudinal subject records across 44 completed ALS clinical trials, where all data available in the PRO-ACT database have been volunteered by PRO-ACT Consortium members. The database includes standardized participant-level information on baseline characteristics, longitudinal clinical outcomes, laboratory measurements, treatment assignment, and survival. Censoring due to death is common in ALS trials, and has received increasing attention given recent emphasis on post-treatment intercurrent events and estimand choice in analyses of functional endpoints \citep{weemering2026heterogeneity}. Relatedly, survivor estimands have been increasingly studied as a way to assess treatment effects on functional outcomes subject to truncation by death in ALS studies \citep{van2022functional,geller2026evaluating,ortholand2026longitudinal}. Motivated by these concerns, we use PRO-ACT to assess how similar the time-varying survivor strata are with respect to commonly collected baseline covariates in ALS studies.

We consider a scenario where the primary study endpoint is the effect of treatment on the ALSFRS-R total score, a standard functional endpoint in ALS studies \citep{cedarbaum1999alsfrs}. We aim to assess the representativeness of commonly collected baseline covariates across the time-varying survivor strata, predictive of disease progression and components of the ENCALS survival model \citep{westeneng2018prognosis}, a commonly-used survival model in ALS research. Specifically, we consider age, sex, baseline ALSFRS-R total score, vital capacity (\% predicted), bulbar onset, riluzole use at baseline, BMI, and the pre-baseline ALSFRS-R progression rate, commonly referred to as $\Delta$FRS \citep{witzel2021neurofilament}. Additional details on all baseline covariates are provided in the Supplementary Material.

Because PRO-ACT masks trial identifiers to preserve privacy, the analysis cohort is pooled across multiple PRO-ACT trials, with information on whether each individual received the active treatment or placebo for their specific trial. In turn, we treat ``Active'' as a heterogeneous label spanning several investigational drugs, and interpret $\hat\tau_t$ descriptively rather than as the effect of a single well-defined treatment. Importantly, within any trial contributing to PRO-ACT, monotonicity implies that individuals surviving under placebo would also survive under that trial's active treatment. In turn, the survivor stratum is still a function of placebo survival under Assumption \ref{as:monotonicity}, meaning $\delta_{kt}$ and $\psi_t$ do not depend on the heterogeneous treatments comprising $A=1$. We provide additional discussion of Assumptions \ref{as:consistency}-\ref{as:monotonicity} in the Supplementary Material.

We focus on the $2082$ participants with complete data on baseline covariates defined in the Supplementary Material, so that all estimates pertain to this complete-covariate cohort. At each month $t$, participants whose survival status could not be determined by $t$ were treated as censored; the proportions censored at $t=3,6,9,12$ months were $4.5\%, 15.0\%, 32.2\%$, and $51.7\%$, respectively. Crude cumulative mortality among all participants through $t=3,6,9,12$ months was $0.6\%, 2.9\%, 6.3\%$, and $9.8\%$, respectively, with a total of $205$ recorded deaths by 12 months among the $2082$ participants.  Throughout, $Y_t$ denotes the ALSFRS-R total score at $t$ months, observed for trial participants who are alive and uncensored at month $t$ and have an ALSFRS-R visit within $\pm 30$ days of $t$.

We estimate $\bm \delta_t$, $\psi_t$, and $\tau_t$ at $t \in \{3,6,9,12\}$ months, using one-step estimators with five-fold cross-fitting which further adjust for censoring due to dropout that we describe in the Supplementary Material, noting these methods are a straightforward extension of those presented in Section \ref{sec:methods}. Both $g_0(\bX,t)$ and the survival-conditional outcome regressions $m_a(\bX,t)$ are estimated with a SuperLearner ensemble \citep{van2007super} of generalized linear models and multivariate adaptive regression splines. Letting $R_t$ indicate whether a participant has known survival status through time $t$, we estimate $\PP(R_t =1 \mid \bX, A)$ with the same SuperLearner ensemble. In estimating the SACE, we additionally account for ALSFRS-R missingness. Since PRO-ACT pools trials with differing Active/Placebo randomization ratios, we estimate the propensity score $e(\bX)$ by logistic regression. Full data preparation and estimation details, and additional details on the above baseline covariates, are included in the Supplementary Material.

\subsection{Results}

\begin{figure}[h!]
    \centering
    \includegraphics[width=0.9\linewidth]{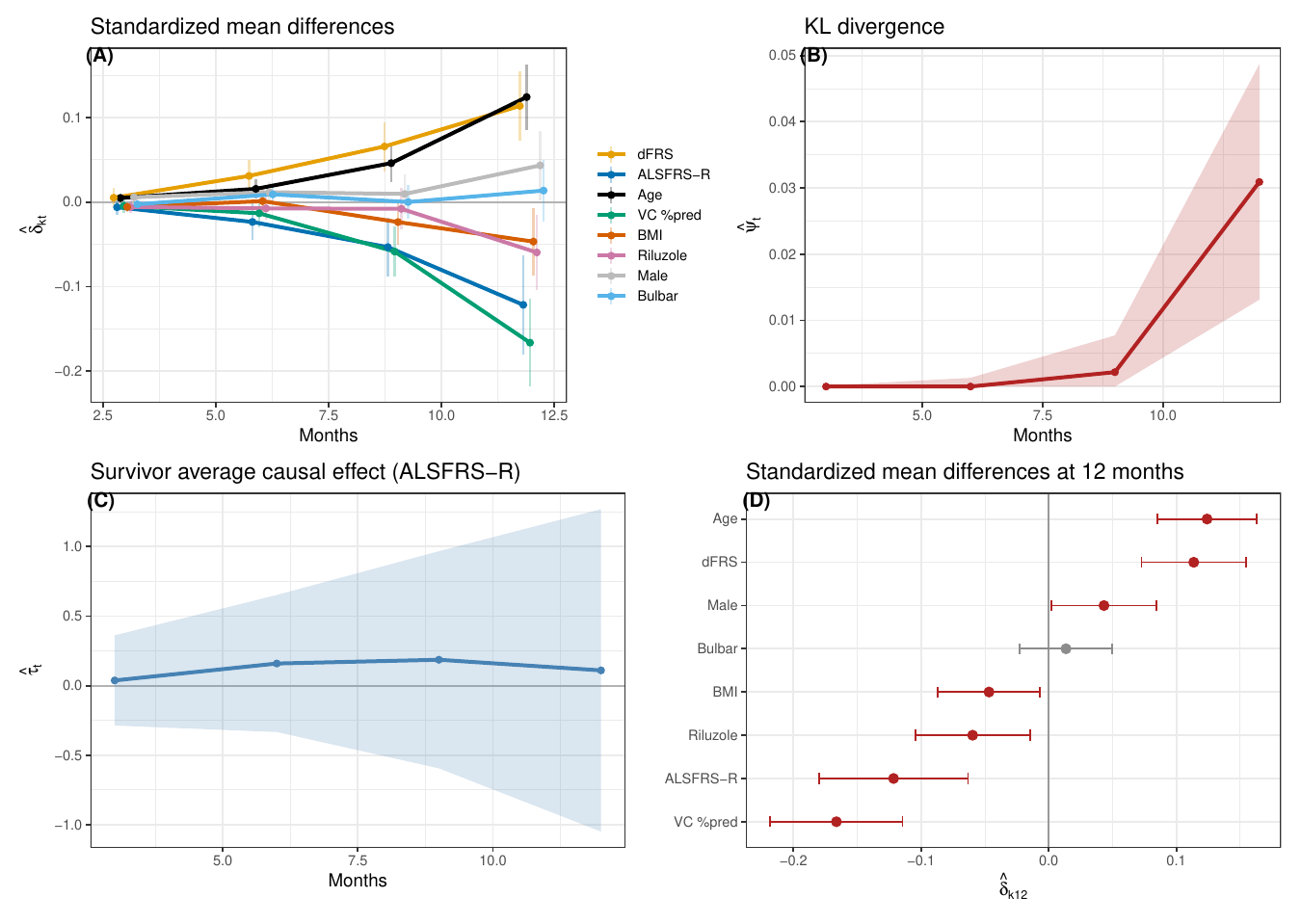}
    \caption{PRO-ACT application results, estimating $\tau_t$, $\bm \delta_t$ and $\psi_t$ at $t \in \{3,6,9,12\}$ months.
    \ifbiostatistics
\\ \textbf{Alt text:} Point estimates from the PRO-ACT application.  The estimated SMDs and KL divergence both increase over follow-up, while SACE point estimates are near-zero with confidence intervals that include zero at all time points.
\fi}
    \label{fig:pro-act-results}
\end{figure}

Figure \ref{fig:pro-act-results} displays the estimated $\bm \delta_t$ and $\psi_t$. Recalling the form of $\delta_{kt}$, positive estimates imply larger baseline means relative to the survivor population.  Early in follow-up, the survivor stratum closely resembles the baseline population. Through $t=6$, no covariate has an estimated SMD exceeding $0.04$ in magnitude, and $\psi_t$ is essentially zero. Although these estimated divergences grow at larger $t$, they remain modest in magnitude throughout the follow-up window. By $t=12$, all but one covariate have SMDs whose $95\%$ confidence intervals exclude zero. Relative to the enrolled population, survivors at $t=12$ are estimated to have higher vital capacity ($\hat \delta_{\text{VC},12}^{\text{OS}} = -0.17$, CI $[-0.21, -0.11]$), younger age ($0.12$, $[0.09, 0.16]$), higher baseline ALSFRS-R ($-0.12$, $[-0.18, -0.06]$), and slower pre-baseline progression ($0.11$, $[0.07, 0.15]$). These estimates are consistent with existing consensus that better respiratory function, younger age, higher baseline function, and slower progression predict longer survival \citep{calvo2020prognostic,labra2016rate,benatar2024prognostic}. A subset of baseline characteristics---including site of symptom onset, riluzole use and BMI---exhibit much smaller estimated divergences at $t=12$. Throughout,  the estimated SACEs $\tau_t$ are near zero at earlier time points and modestly positive at later time points, with 95\% confidence intervals including zero throughout, consistent with limited evidence of functional benefit for many existing ALS therapies.

While a subset of estimated $\delta_{kt}$ are distinguishable from 0 by $t=12$, none of these estimates exceed 0.2 in magnitude, a threshold typically interpreted as moderate imbalance in the matching and weighting literatures \citep{austin2009balance}. The estimated KL divergence at $t=12$ is similarly small, rising from near-zero  to  $0.031$ by $t=12$.  Practically, these results suggest that in terms of the above set of baseline covariates, survivors are broadly similar to the original baseline population over shorter follow-up horizons, and only modestly different over longer ones. 

\begin{figure}[t!]
\centering
\includegraphics[width=0.95\linewidth]{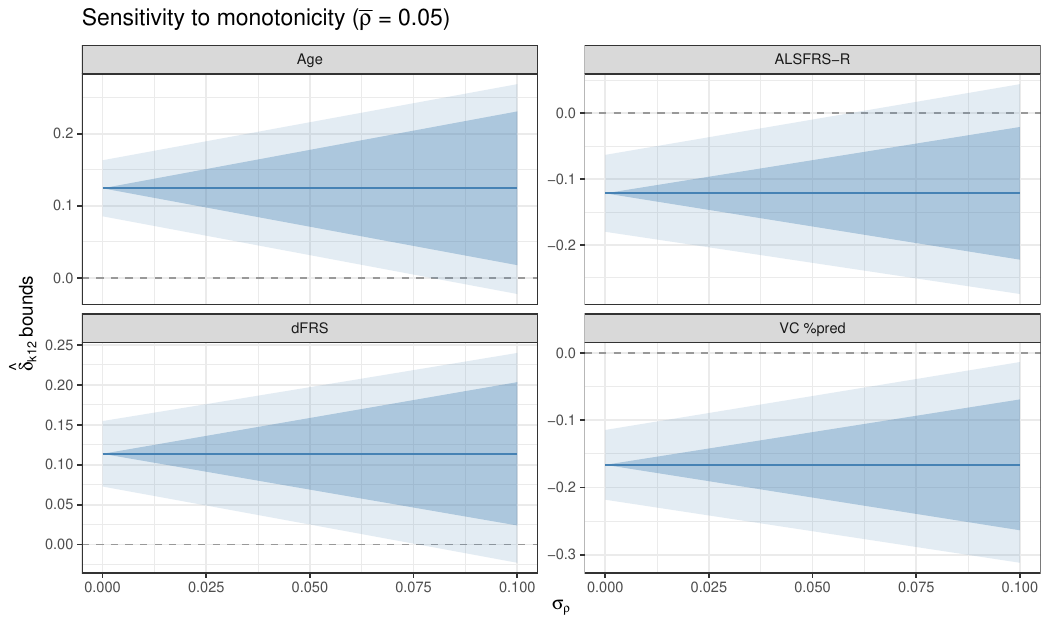}
\caption{Sensitivity analysis for $\hat \delta_{k12}^{\text{OS}}$ in the PRO-ACT cohort at $t =12$ months, fixing $\bar \rho_{12} = 0.05$. Inner dark blue band denotes point sensitivity bounds, and outer light blue band denotes 95\% CIs which account for uncertainty in bound estimation.
\ifbiostatistics
\\ \textbf{Alt text:} Estimated sensitivity bounds at 12 months for age, ALSFRS-R, progression rate, and vital capacity. The estimated point bounds and their associated confidence intervals widen as the sensitivity parameter increases.
\fi}
\label{fig:sens-bounds}
\end{figure}

\paragraph{Sensitivity to monotonicity.} Using the sensitivity analysis framework of Section \ref{sec:relaxing-monotonicity}, Figure \ref{fig:sens-bounds} presents estimated SMD bounds at $t=12$ for the four covariates with the largest $|\hat \delta_{k,12}^{\text{OS}}|$---age, vital capacity, pre-baseline progression rate, and baseline ALSFRS-R---across $\sigma_\rho \in [0, 0.10]$, fixing $\bar \rho_t = 0.05$. The estimated point bounds for all four covariates remain entirely on one side of zero across this range, excluding zero through $\sigma_\rho = 0.1$. Conversely, the estimated 95\% CIs that account for uncertainty in the point bounds include 0 at larger values of $\sigma_{\rho_t}$ for all covariates excluding percent-predicted VC. These results suggest that the conclusion that the survivor stratum differs from the baseline population in these clinical features is robust to modest covariate-dependent variation in $\rho(\bX,t)$.

\section{Discussion}
\label{sec:discussion}

Clinical studies are increasingly incorporating endpoints based on survivor principal strata to accommodate censoring due to death, but it can be difficult to determine whether these strata are representative of the baseline trial populations that treatments are intended for. In this paper, we've proposed complementary time-indexed divergence measures that quantify how similar the survivor stratum is to the baseline population, and how that similarity changes over follow-up. These estimands are identified under a subset of the assumptions required to identify the time-indexed SACE, and our proposed one-step estimators require only  a subset of the nuisance functions already estimated in standard SACE analyses. We  demonstrate that estimation of the KL divergence $\psi_t$---and a broader set of scalar distributional distance measures---is materially harder than estimation of covariate SMDs $\delta_{kt}$; in randomized settings where the treatment assignment mechanism is known, estimation of $\psi_t$ still requires consistent and sufficiently fast estimation of the placebo survival function $g_0(\bX,t)$, whereas consistent and asymptotically normal estimation of $\delta_{kt}$ is guaranteed under arbitrary misspecification of this function. We additionally show that violations to monotonicity induce bias only to the extent that these violations vary in $\bX$, and we in turn provide a pragmatic sensitivity analysis framework that only requires specifying the mean and variance of the standard monotonicity sensitivity function.

There are numerous directions for future work. Our proposed divergence measures weight all covariates equally, while in practice a given divergence may be considered more consequential for covariates that strongly predict the functional outcome of interest. Outcome-informed divergence measures which more heavily weight covariates predictive of the functional outcome would provide a possibly more informative diagnostic tool for researchers. Relatedly, the sensitivity bounds presented in Section \ref{sec:relaxing-monotonicity} are conservative by construction. Sharper partial identification methods that leverage additional structure would be a valuable contribution. Finally, our efficiency theory and resulting debiased estimators arise from models that impose no structure on the distribution of functional outcomes or survival over time. A promising direction is to construct debiased estimators with improved efficiency by imposing restrictions on functional and survival outcome distributions \citep{stephens2014locally}, and to account for broader and more complex dropout mechanisms. For the standardized mean differences, data-adaptive shrinkage approaches like those outlined in \citep{susmann2026asymptotically} could be used to improve finite-sample efficiency, as well as approaches from the empirical efficiency maximization literature \citep{rubin2008empirical}.

\clearpage 
\subsection*{PRO-ACT Consortium Attribution}
\label{sec:proact-consortium}

\noindent$^*$Data used in the preparation of this article were obtained from the Pooled Resource Open-Access ALS Clinical Trials (PRO-ACT) Database. As such, the following organizations and individuals within the PRO-ACT Consortium contributed to the design and implementation of the PRO-ACT Database and/or provided data, but did not participate in the analysis of the data or the writing of this report: Alexion Pharmaceuticals, Inc.; ALS Therapy Alliance; Amylyx Pharmaceuticals, Inc.; Apellis Pharmaceuticals, Inc.; Cytokinetics, Inc.; Knopp Biosciences; Neuraltus Pharmaceuticals, Inc.; Neurological Clinical Research Institute, MGH; Northeast ALS Consortium; Novartis; Orion Corporation; Prize4Life Israel; PTC Therapeutics, Inc.; Regeneron Pharmaceuticals, Inc.; Sanofi; Teva Pharmaceutical Industries, Ltd.; The ALS Association; The Sean M.\ Healey \& AMG Center for ALS at Massachusetts General Hospital; and University Medical Center Utrecht.

\ifbiostatistics
    % if using biostats format all this stuff already appears at start
  \else

\paragraph{Acknowledgements} We thank Sophie Woodward and Eric Macklin for helpful comments and suggestions.
\\[.8em]
\textbf{Funding:} This work was supported in part by National Institutes of Health Grant R01DA056407.
\\[.8em]
\textbf{Conflicts of interest:} W.B. and J.G. are current and former employees of the Neurological Clinical Research Institute (NCRI) at Mass General Brigham, respectively. K.B. previously received doctoral funding support from the NCRI. The NCRI manages the PRO-ACT database used in this study but had no role in the conduct or reporting of the work. The authors declare no other conflicts of interest relevant to this work.
\\[.8em]
\textbf{Data and code availability:} The data analyzed in this study were obtained from the PRO-ACT database. Patient-level PRO-ACT data are available to registered users under the PRO-ACT Terms and Conditions and cannot be redistributed by the authors. Aggregate results and analysis code may be shared in accordance with those terms. Access to the database may be requested at \url{https://ncri1.partners.org/ProACT/}. Code to replicate the simulation exercises in Section \ref{sec:simulation} is available at \url{https://github.com/keithbarnatchez/survstrat-paper}, and an R package for implementing the proposed methods is available at \url{https://github.com/keithbarnatchez/survstrat}.

\fi

\bibliographystyle{apalike}
\bibliography{sources}

\clearpage

\appendix
\setcounter{page}{1}

\begin{center}
    \Large{\textbf{Characterizing Survivor Principal Strata Over Time: Identification, Efficient Estimation, and Sensitivity Analysis}
    \\
    Supplementary Materials}
    \\[1em]
    Keith Barnatchez$^{1,\dagger}$, Willow Butler$^{2}$, Julia A. Geller$^{2,3}$, Elizabeth A. Stuart$^{1}$, \\[-0.5em]
{\small The Pooled Resource Open-Access ALS Clinical Trials Consortium$^*$} \\[0.1em]
\small
$^{1}$Department of Biostatistics, Johns Hopkins Bloomberg School of Public Health \\
\small $^{2}$Neurological Clinical Research Institute, Massachusetts General Hospital \\
\small $^{3}$Computer Science and Engineering Division, University of Michigan \\[1em]
\noindent\textbf{$^\dagger$Corresponding author.} Keith Barnatchez, Department of Biostatistics,
Johns Hopkins Bloomberg School of Public Health, 615 N Wolfe St, Baltimore, MD 21205, USA.
Email: \href{mailto:kbarnat1@jh.edu}{kbarnat1@jh.edu}
\end{center}

\section{Proofs}

\paragraph{Proof of Theorem \ref{thm:theta-id}}
First notice
\begin{align*}
    \E[\bX \mid S_t(0)=S_t(1)=1] &= \E[\bX \mid S_t(0)=1] \\
    &= \E\left[ \frac{\bX \cdot \mathbb{I}(S_t(0)=1)}{\mathbb{P}(S_t(0)=1)} \right] \\
    &= \E\left[ \frac{\bX \cdot \mathbb{I}(S_t(0)=1)}{\E[\PP(S_t=1 \mid A=0,\bX)]} \right],
\end{align*}
since $\mathbb{P}(S_t(0)=1) = \E[\PP(S_t=1 \mid A=0,\bX)]=\mu_{0t}$ under Assumptions \ref{as:consistency} and \ref{as:surv-unc}. Then, focusing on the numerator, notice
\begin{align*}
    \E[\bX \mathbb{I}(S_t(0)=1)] &= \E[\bX  \PP(S_t(0)=1 \mid \bX) ] \\
    &=  \E[\bX  \PP(S_t=1 \mid \bX,A=0)] \\
    &= \E[\bX  \cdot g_0(\bX,t)] 
\end{align*}
Noting the denominator $\E[\PP(S_t=1 \mid A=0,\bX)] = \E[g_0(\bX,t)]$ yields the desired result.

\paragraph{Proof of Theorem \ref{thm:kl-id}} We follow similar arguments used to prove Theorem \ref{thm:theta-id}. First, notice
\[
\psi_t = \int p(\bx \mid S_t(0)=S_t(1)=1) \log \frac{p(\bx \mid S_t(0)=S_t(1)=1)}{p(\bx)}  d\bx.
\]
Under Assumption \ref{as:monotonicity}, $p(\bx \mid S_t(0)=S_t(1)=1) = p(\bx \mid S_t(0)=1)$. By Bayes' rule, we have
\[
\frac{p(\bx \mid S_t(0)=1)}{p(\bx)} = \frac{\PP(S_t(0)=1 \mid \bX = \bx)}{\PP(S_t(0)=1)} = \frac{g_0(\bx,t)}{\mu_{0t}} = w_0(\bx,t),
\]
where the second equality uses Assumptions \ref{as:consistency} and \ref{as:surv-unc} as in the proof of Theorem \ref{thm:theta-id}. Noting $p(\bx \mid S_t(0)=1) = w_0(\bx,t) p(\bx)$, it follows that
\[ 
\psi_t = \int w_0(\bx,t) \log w_0(\bx,t) p(\bx) d\bx = \E[w_0(\bX,t) \cdot \log w_0(\bX,t)].
\]

\paragraph{Proof of Theorem \ref{thm:eif-thm}} To simplify presentation and calculations throughout, we follow \cite{kennedy2024semiparametric} and temporarily assume the observed data are completely discrete, letting the operator $\text{IF}(\cdot)$ map functionals to their corresponding influence function in a discrete, fully nonparametric statistical model. 
\\\\
We first focus on $\theta_{kt}$, where we recall $\theta_{kt} = N_{kt}/D_t$, implying we can first obtain the EIFs of $N_{kt} \eqdef \E[X_k g_0(\bX,t)]$ and $D_t \eqdef \mu_{0t} = \E[g_0(\bX,t)]$ and then apply the delta method. Noting $N_{kt}$ and $D_t$ are both linear functionals, it is straightforward to verify \citep{kennedy2024semiparametric}
\begin{align*}
\varphi_{N_{kt}}(\bm O) &= \frac{\II(A=0)}{1-e(\bX)}\{S_t - g_0(\bX,t)\} X_k + g_0(\bX,t) X_k - N_{kt}, \\
\varphi_{D_t}(\bm O) &= \frac{\II(A=0)}{1-e(\bX)}\{S_t - g_0(\bX,t)\} + g_0(\bX,t) - D_t.
\end{align*}
Then, the delta method implies
\begin{align*}
    \varphi_{\theta_{k,t}}(\bm O) &= \frac{1}{D_t}\left[\varphi_{N_{kt}}(\bm O) - \theta_{kt} \varphi_{D_t}(\bm O)\right] \\ 
    &= \frac{1}{\mu_{0t}}\left[\frac{\II(A=0)}{1-e(\bX)}\{S_t - g_0(\bX,t)\}(X_k - \theta_{kt}) + g_0(\bX,t)(X_k - \theta_{kt})\right],
\end{align*}
where the final line holds since $N_{kt} - \theta_{kt} D_t = 0$.
\\\\
For $\psi_t = \E[w_0(\bX,t) \log w_0(\bX,t)]$, we begin by noting
\[
\text{IF}\{w_0(\bx,t)\} = \frac{\text{IF}\{g_0(\bx,t)\}}{\mu_{0t}} - \frac{g_0(\bx,t)}{\mu_{0t}^2}\text{IF}\{\mu_{0t}\}.
\]
Then, letting $h(w) = w\log w$, we have that
\begin{align*}
    \text{IF}(\psi_t) &= \sum_{\bx} h\{w_0(\bx,t)\} \text{IF}\{p(\bx)\} + \sum_{\bx} p(\bx) h'\{w_0(\bx,t)\} \text{IF}\{w_0(\bx,t)\}
\end{align*}
Noting $h'(w) = \log w + 1$ and substituting yields
\[
\varphi_{\psi_t}(\bm O) = \frac{\II(A=0)}{1-e(\bX)} \cdot \frac{\log w_0(\bX,t) - \psi_t}{\mu_{0t}} \cdot \{S_t - g_0(\bX,t)\} + w_0(\bX,t)\{\log w_0(\bX,t) - \psi_t - 1\} + 1,
\]
since $\E[w_0(\bX,t)]=1$ and $\E[w_0(\bX,t)\{\log w_0(\bX,t) +1 \}] = \psi_t +1$. Since we consider the fully unrestricted model for the observed data, the tangent space is all of $L^2_0(\mathbb{P})$, implying both $\varphi_{\psi_t}(\bm O)$ and $\varphi_{\theta_{kt}}(\bm O)$ are members of the tangent space and in turn the EIFs for their respective functionals.

\paragraph{Proof of Theorem \ref{thm:theta-asym}}

We first show that, under the conditions of Theorem \ref{thm:theta-asym}, $\hat \theta_{kt}^\text{OS}$ is consistent if one of $\hat g_0(\bX,t)$ or $\hat e(\bX)$ is consistent.
\\ \\
To start, recall $\hat \theta_{kt}^{\text{OS}} = \hat N_{kt}^\text{OS}/\hat D_t^\text{OS}$. Letting $g^* = \operatorname{plim} \hat g_0$ and $e^* = \operatorname{plim} \hat e$, and focusing on 
\begin{align*}
\hat N_{kt}^\text{OS} &\xrightarrow{p} \E[g^*(\bX,t) X_k] + \E\left[\frac{\II(A=0)}{1-e^*(\bX)}\{S_t - g^*(\bX,t)\} X_k\right] \\
            &= \E[g^*(\bX,t) X_k] + \E\left[\frac{1-e(\bX)}{1-e^*(\bX)}\{g_0(\bX,t) - g^*(\bX,t)\} X_k\right]
\end{align*}
by iterated expectations. If $g^* = g_0$, the second term is zero and $\hat N_{kt} \to \E[g_0 X_k] = N_{kt}$. Similarly, if $e^* = e$, the ratio $(1-e)/(1-e^*)$ converges to 1 and the same conclusion holds. An analogous argument for $\hat D_t^\text{OS}$ gives $\hat D_t^\text{OS} \to \mu_{0t}$ under either condition, and Slutsky's theorem yields $\hat \theta_{kt}^{\text{OS}} \to \theta_{kt}$.
\\ \\
To establish asymptotic linearity, let $\eta = (g_0, e)$ and $\hat \eta = (\hat g_0, \hat e)$ collect the true and estimated nuisance functions. 
For the ratio estimator in Equation \eqref{eq:theta-onestep-ratio}, we can write
\[
\hat\theta_{kt}^{\mathrm{OS}}-\theta_{kt}
=\frac{1}{\hat D_t^{\mathrm{OS}}}\E_n\left[(X_k-\theta_{kt})
\left\{\hat g_0(\bX,t)+\frac{\II(A=0)}{1-\hat e(\bX)}
\{S_t-\hat g_0(\bX,t)\}\right\}\right].
\]
Adding and subtracting the above at the true and estimated $\eta$, after rearranging and by iterated expectations we arrive at the following expansion:
\begin{align*}
\hat\theta_{kt}^{\mathrm{OS}}-\theta_{kt}
=&\E_n\{\varphi_{\theta_{kt}}(\bm O)\}\\
&+\left(\frac{\mu_{0t}}{\hat D_t^{\mathrm{OS}}}-1\right)
\E_n\{\varphi_{\theta_{kt}}(\bm O)\}\\
&+\frac{1}{\hat D_t^{\mathrm{OS}}}(\E_n-\E)
\Big[(X_k-\theta_{kt})\Big\{
\hat g_0(\bX,t)+\frac{\II(A=0)}{1-\hat e(\bX)}
\{S_t-\hat g_0(\bX,t)\}\\[-0.25em]
&\hspace{5.5cm}
-g_0(\bX,t)-\frac{\II(A=0)}{1-e(\bX)}
\{S_t-g_0(\bX,t)\}\Big\}\Big]\\
&+\frac{1}{\hat D_t^{\mathrm{OS}}}
\E\left[(X_k-\theta_{kt})
\frac{e(\bX)-\hat e(\bX)}{1-\hat e(\bX)}
\{\hat g_0(\bX,t)-g_0(\bX,t)\}\right] \\
&= \text{I} + \text{II} +  \text{III} +  \text{IV}.
\end{align*}
For term I, notice $\sqrt n \E_n\{\varphi_{\theta_{kt}}(\bm O)\} \xrightarrow{d} \mathcal{N}(0, \E[\varphi_{\theta_{kt}}(\bm O)^2])$ by the Central Limit Theorem and since $\E[\varphi_{\theta_{kt}}(\bm O)]=0$.  Since $\hat D_t^\text{OS} \xrightarrow{p}{\mu_{0t}}$ so long as one of $g_0$ or $e$ is consistent, term II is $o_\PP(1) \cdot O_\PP(1/\sqrt n) = o_\PP(1/\sqrt n)$. Term III is an empirical process term, which is $o_\PP(1/\sqrt n)$ provided $\hat g_0$ and $\hat e$ are obtained from a separate sample, which is ensured by cross-fitting \citep{kennedy2024semiparametric}. Finally, term IV is $O_\PP(||\hat e - e||_2 \cdot ||\hat g_0 - g_0 ||_2 )$ by Cauchy-Schwarz. Combining these four terms implies 
\[
\hat \theta_{kt}^\text{OS} - \theta_{kt} = \E_n\{\varphi_{\theta_{kt}}(\bm O)\} + o_\PP(1/\sqrt n) + O_\PP(||\hat e - e||_2 \cdot ||\hat g_0 - g_0 ||_2 ),
\]
as desired.

\paragraph{Proof of Theorem \ref{thm:psi-asym}}

We use an analogous argument to the one used in the proof of Theorem \ref{thm:theta-asym} above, with one additional contribution to $R_{n,\psi_{kt}}$ due to the nonlinear dependence of $\psi_t$ on $w_0$. Analogous to the proof of Theorem \ref{thm:theta-asym}, we consider the following expansion of $\hat \psi_t^{\text{OS}}$:
\begin{align*}
\hat\psi_t^{\mathrm{OS}}-\psi_t
=&\E_n\{\varphi_{\psi_t}(\bm O)\}\\
&+(\E_n-\E)
\{\hat\varphi_{\psi_t}(\bm O)-\varphi_{\psi_t}(\bm O)\}\\
&+\hat\psi_t^{\mathrm{PI}}-\psi_t
+\E\{\hat\varphi_{\psi_t}(\bm O)\} \\
&=  \text{I} + \text{II} +  \text{III}
\end{align*}
Above, for Term I note $\sqrt n \E_n\{\varphi_{\psi_t}(\bm O)\} \xrightarrow{d} \mathcal{N}(0, \E[\varphi_{\psi_t}(\bm O)^2])$ by the Central Limit Theorem. Under cross-fitting,  term II is $o_\PP(1/\sqrt n)$. Throughout, to compress notation and to demonstrate results for similar functionals discussed in Appendix \ref{app:alt-divergences}, we let $h(w_0(\bX,t)) = w_0(\bX,t) \log w_0(\bX,t)$.
Expanding $\E\{\hat\varphi_{\psi_t}(\bm O)\}$, term III can be rewritten as
\begin{align*}
&\hat\psi_t^{\mathrm{PI}}-\psi_t
+\E\{\hat\varphi_{\psi_t}(\bm O)\}\\
=&-\E\left[h\{w_0(\bX,t)\}-h\{\hat w_0(\bX,t)\}-h'\{\hat w_0(\bX,t)\}\{w_0(\bX,t)-\hat w_0(\bX,t)\}\right]\\
&+\left(\frac{\mu_{0t}}{\hat\mu_{0t}}-1\right)\E\left[h'\{\hat w_0(\bX,t)\}\{w_0(\bX,t)-\hat w_0(\bX,t)\}\right]\\
&+\frac{1}{\hat\mu_{0t}}\E\Big[\frac{\hat e(\bX)-e(\bX)}{1-\hat e(\bX)}
\{g_0(\bX,t)-\hat g_0(\bX,t)\}\times\Big\{\log\hat w_0(\bX,t)-\E[\hat w_0(\bX,t)\log\hat w_0(\bX,t)]\Big\}\Big] \\
&= \text{A} + \text{B} + \text{C}
\end{align*}
Above, term B is $O_\PP(||\hat g_0 - g_0 ||_2^2 + 1/\sqrt n )$, since $\hat \mu_{0t} - \mu_{0t} = O_\PP(||\hat g_0 - g_0||_2 + 1/\sqrt n)$ and $\hat w_0 - w_0 = O_\PP(||\hat g_0 - g_0||_2 + 1/\sqrt n)$. Term C is $O_\PP(||\hat e - e ||_2 \cdot ||\hat g_0 - g_0 ||_2 )$ by Cauchy-Schwarz and the positivity Assumption \ref{as:positivity}.
\\\\
Finally, by taking a second-order Taylor approximation of $w_0(\bx,t)$, notice that term A above can be written
\begin{align*}
\E\left[h\{w_0(\bX,t)\}-h\{\hat w_0(\bX,t)\}
-h'\{\hat w_0(\bX,t)\}\{w_0(\bX,t)-\hat w_0(\bX,t)\}\right] &= \E\left[ \frac{\{\hat w_0(\bX,t)-w_0(\bX,t)\}^2}{2\tilde w(\bX,t)}\right] \\
&= O_\PP(||\hat g_0 - g_0 ||_2^2)
\end{align*}
for some $\tilde w(\bx,t)$ that lies between  $\hat w_0(\bx,t)$ and $w_0(\bx,t)$ for all $\bx$ with positive support. Recalling the forms of terms A and B, as well as terms I and II above, we have that
\begin{align*}
    \hat\psi_t^{\mathrm{OS}}-\psi_t &= \E_n\{\varphi_{\psi_t}(\bm O) \} + o_\PP(1/\sqrt n)  +  O_\PP\{ \ ||\hat g_0 - g_0 ||_2^2+||\hat e - e ||_2 \cdot ||\hat g_0 - g_0 ||_2  \}.
\end{align*}

\paragraph{Proof of Proposition \ref{prop:work-mod-asym} }

 With $\hat e(\bX) = e(\bX)$, recall that  the second-order remainder $R_{n,\theta_{kt}}$ from the proof of Theorem \ref{thm:theta-asym} is $O_\PP(\|\hat g_0 - g_0\|_2 \cdot \|\hat e - e\|_2)=0$. Letting $g^* = \operatorname{plim} \hat g_0$, we have
\begin{align*}
\E\{\varphi_{\theta_{kt}}^*(\bm O)\}
&=\frac{1}{\mu_{0t}}\E\left[
\{g_0(\bX,t)-g^*(\bX,t)\}(X_k-\theta_{kt})
+g^*(\bX,t)(X_k-\theta_{kt})\right]\\
&=\frac{1}{\mu_{0t}}\E\{g_0(\bX,t)(X_k-\theta_{kt})\}=0,
\end{align*}
regardless of whether $g^* = g_0$. The same arguments from the proof Theorem \ref{thm:theta-asym} then imply $\hat \theta_{kt}^\text{OS}$ is asymptotically linear for $\theta_{kt}$.

\paragraph{Proof of Proposition \ref{prop:const-mono}}

When $\rho(\bX,t) = \bar \rho_t$ is constant in $\bX$, $(1 - \bar \rho_t)$  factors out of both the numerator and the denominator of the survivor identifying functional $\theta_{kt}$. To see this, notice
\[
\theta_{kt} = \frac{\E[X_k \cdot \PP(S_t(0)=S_t(1)=1 \mid \bX)]}{\PP(S_t(0)=S_t(1)=1)} = \frac{\E[X_k \cdot (1 - \rho(\bX,t)) g_0(\bX,t)]}{\E[(1 - \rho(\bX,t)) g_0(\bX,t)]},
\]
where the second equality uses $\PP(S_t(0)=S_t(1)=1 \mid \bX) = \PP(S_t(0)=1 \mid \bX) - \PP(S_t(0)=1, S_t(1)=0 \mid \bX) = (1 - \rho(\bX,t)) g_0(\bX,t)$. Substituting $\rho(\bX,t) = \bar \rho_t$ gives
\[
\theta_{kt} = \frac{(1-\bar \rho_t) \E[X_k g_0(\bX,t)]}{(1-\bar \rho_t) \E[g_0(\bX,t)]} = \frac{\E[X_k g_0(\bX,t)]}{\E[g_0(\bX,t)]} = \theta_{kt}^{\text{mono}}
\]
An analogous argument implies $\bm \delta_t$ and $\psi_t$ remain identified under constant violations to monotonicity.

\paragraph{Proof of Proposition \ref{prop:sens-bias}}

We begin by noting
\begin{align*}
\theta_{kt}^{\mathrm{true}}
&=\frac{\E[X_k\{1-\rho(\bX,t)\}\mid S_t(0)=1]}
{\E[1-\rho(\bX,t)\mid S_t(0)=1]}\\
&=\frac{\E[X_k\{1-\rho(\bX,t)\}\mid S_t(0)=1]}
{1-\bar\rho_t},
\end{align*}
whereas $\theta_{kt}^{\mathrm{mono}}=\E[X_k\mid S_t(0)=1]$. Then, notice
\begin{align*}
  \theta_{kt}^{\text{true}} - \theta_{kt}^{\text{mono}} &=  \frac{\E(X_k\mid S_t(0)=1) - \E[X_k \cdot \rho(\bX,t)\mid S_t(0)=1]}{1-\bar \rho_t} - \E(X_k\mid S_t(0)=1)
   \\
   &=-\frac{\Cov(\rho(\bX,t), X_k | S_t(0)=1)}{1 - \bar \rho_t}.
\end{align*}
Then, Cauchy--Schwarz implies
\[
|\Cov(\rho(\bX,t), X_k| S_t(0)=1)| \leq \sigma_{\rho_t} \cdot \text{SD}(X_k \mid S_t(0) = 1),
\]
Substituting into the above and rearranging yields
\[
\big|\theta_{kt}^{\text{true}} - \theta_{kt}^{\text{mono}}\big| \leq \frac{\sigma_{\rho_t}  \cdot \text{SD}(X_k \mid S_t(0)=1)}{1 - \bar \rho_t},
\]
as desired.
\section{Construction of  doubly robust sensitivity bounds}
\label{app:sens-bounds}

\subsection{Standardized mean differences}

\paragraph{One-step estimator for $s_{kt} = \text{SD}(X_k\mid S_t(0)=1)$}
Under monotonicity, $s_{kt}^2 = \nu_{kt} - \theta_{kt}^2$, where
\[
\nu_{kt} = \E[X_k^2 \mid S_t(0)=1] = \frac{\E[X_k^2  g_0(\bX,t)]}{\E[g_0(\bX,t)]}.
\]
Because $\nu_{kt}$ shares the ratio-of-expectations identifying functional of $\theta_{kt}$ with $X_k$ replaced by $X_k^2$, its EIF and one-step
estimator have similar forms (Section \ref{sec:methods}):
\[
\varphi_{\nu_{kt}}(\bm O) = \frac{X_k^2 - \nu_{kt}}{\mu_{0t}}\left[g_0(\bX,t) + \frac{\mathbb{I}(A=0)}{1-e(\bX)}\{S_t - g_0(\bX,t)\}\right],
\]
\[
\hat \nu_{kt}^{\text{OS}} = \hat M_{kt}/\hat D_t, \qquad \hat M_{kt} = \sumn \left[\hat g_0(\bX_i,t) X_{ik}^2 + \frac{\mathbb{I}(A_i=0)}{1-\hat
e(\bX_i)}\{S_{i,t} - \hat g_0(\bX_i,t)\} X_{ik}^2\right],
\]
where $\hat D_t$ is the AIPW estimator of $\mu_{0t}$ from Section \ref{sec:methods}. Combining $\hat \nu_{kt}^{\text{OS}}$ and $\hat \theta_{kt}
^{\text{OS}}$ gives
\[
\varphi_{s_{kt}}(\bm O)
=\frac{\varphi_{\nu_{kt}}(\bm O)-2\theta_{kt}\varphi_{\theta_{kt}}(\bm O)}
{2s_{kt}},
\]
yielding
\[
\hat s_{kt}^{\text{OS}} = \sqrt{\max\big(\hat \nu_{kt}^{\text{OS}} - (\hat \theta_{kt}^{\text{OS}})^2, 0\big)}, \qquad \widehat{\text{SE}}(\hat
s_{kt}^{\text{OS}}) = n^{-1/2}\widehat{\text{sd}}(\hat \varphi_{s_{kt}}).
\]

\paragraph{Bounds on $\theta_{kt}$ and $\delta_{kt}$.}
Plugging $\hat s_{kt}^{\text{OS}}$ into Equation \eqref{eq:sens-bound} of Section \ref{sec:relaxing-monotonicity} gives
\[
\hat \theta_{kt}^{\text{L-OS}} = \hat \theta_{kt}^{\text{OS}} - \frac{\sigma_{\rho_t}}{1-\bar \rho_t}\hat s_{kt}^{\text{OS}}, \qquad \hat
\theta_{kt}^{\text{U-OS}} = \hat \theta_{kt}^{\text{OS}} + \frac{\sigma_{\rho_t}}{1-\bar \rho_t}\hat s_{kt}^{\text{OS}}.
\]
Translating to the SMD scale $\delta_{kt} = (\E[X_k] - \theta_{kt})/\sigma_k$, where $\sigma_k=\mathrm{SD}(X_k)$, gives
\[
\hat \delta_{kt}^{\text{L-OS}} = \hat \delta_{kt}^{\text{OS}} - \frac{\sigma_{\rho_t}}{1-\bar \rho_t}\frac{\hat s_{kt}^{\text{OS}}}{\hat
\sigma_k}, \qquad \hat \delta_{kt}^{\text{U-OS}} = \hat \delta_{kt}^{\text{OS}} + \frac{\sigma_{\rho_t}}{1-\bar \rho_t}\frac{\hat s_{kt}
^{\text{OS}}}{\hat \sigma_k},
\]
where $\hat\sigma_k=\widehat{\mathrm{SD}}_k$ is the empirical standard deviation of $X_k$ defined in Section \ref{sec:methods}.

\paragraph{Confidence interval.}
We can additionally construct an overall confidence interval for $\delta_{kt}$ that reflects the estimation uncertainty in the point bounds. Let $
\mu_k=\E(X_k)$. The influence functions for $s_{kt}$ and $\sigma_k$ are
\[
\varphi_{s_{kt}}(\bm O)
=
\frac{\varphi_{\nu_{kt}}(\bm O)-2\theta_{kt}\varphi_{\theta_{kt}}(\bm O)}{2s_{kt}},
\qquad
\varphi_{\sigma_k}(\bm O)
=
\frac{(X_k-\mu_k)^2-\sigma_k^2}{2\sigma_k}.
\]
The influence function for $\delta_{kt}$ is
\[
\varphi_{\delta_{kt}}(\bm O)
=\frac{1}{\sigma_k}
\left[
(X_k-\mu_k)
-\varphi_{\theta_{kt}}(\bm O)
-\delta_{kt}\frac{(X_k-\mu_k)^2-\sigma_k^2}{2\sigma_k}
\right].
\]
Since
\[
\delta_{kt}^{L}
=
\delta_{kt}
-\frac{\sigma_{\rho_t}}{1-\bar\rho_t}\frac{s_{kt}}{\sigma_k},
\qquad
\delta_{kt}^{U}
=
\delta_{kt}
+\frac{\sigma_{\rho_t}}{1-\bar\rho_t}\frac{s_{kt}}{\sigma_k},
\]
the influence functions for the two bounds are
\[
\varphi_{\delta_{kt}^{L}}(\bm O)
=
\varphi_{\delta_{kt}}(\bm O)-\frac{\sigma_{\rho_t}}{1-\bar\rho_t}
\left\{
\frac{\varphi_{s_{kt}}(\bm O)}{\sigma_k}-\frac{s_{kt}}{\sigma_k^2}\varphi_{\sigma_k}(\bm O)
\right\},
\]
and
\[
\varphi_{\delta_{kt}^{U}}(\bm O)
=
\varphi_{\delta_{kt}}(\bm O)
+\frac{\sigma_{\rho_t}}{1-\bar\rho_t}
\left\{
\frac{\varphi_{s_{kt}}(\bm O)}{\sigma_k}
-\frac{s_{kt}}{\sigma_k^2}\varphi_{\sigma_k}(\bm O)
\right\}.
\]
The reported 95\% confidence band is
\[
\left[
\hat\delta_{kt}^{\mathrm{L-OS}}
-1.96\widehat{\mathrm{SE}}(\hat\delta_{kt}^{\mathrm{L-OS}}),\
\hat\delta_{kt}^{\mathrm{U-OS}}
+1.96\widehat{\mathrm{SE}}(\hat\delta_{kt}^{\mathrm{U-OS}})
\right],
\]
where, for $j\in\{L,U\}$,
\[
\widehat{\mathrm{SE}}(\hat\delta_{kt}^{j\text{-OS}})
=
\frac{1}{\sqrt n}
\left[
\frac{1}{n}\sum_{i=1}^n
\left\{
\hat\varphi_{\delta_{kt}^{j}}(\bm O_i)
-\bar{\hat\varphi}_{\delta_{kt}^{j}}
\right\}^2
\right]^{1/2},
\qquad
\bar{\hat\varphi}_{\delta_{kt}^{j}}
=
\frac{1}{n}\sum_{i=1}^n
\hat\varphi_{\delta_{kt}^{j}}(\bm O_i).
\]

\subsection{Sensitivity Bounds for KL Divergence}

We can similarly construct sensitivity bounds for $\psi_t$ under violations to Assumption \ref{as:monotonicity}. Continuing to let $\rho(\bX,t) = \PP(S_t(1)=0\mid S_t(0)=1,\bX)$, $\bar\rho_t  = \PP(S_t(1)=0\mid S_t(0)=1)$ and $\sigma_{\rho_t} = \text{SD}(\rho(\bX,t) \mid S_t(0)=1)$, we let
\[
\psi_t^\text{mono} = \E[w_0(\bX,t) \log w_0(\bX,t)]
\]
be the identifying functional for $\psi_t$ that is valid under Assumptions \ref{as:consistency}-\ref{as:monotonicity}. Then, noting the true KL-divergence can be written
\begin{align*}
    \psi_t &= 
    \int p(\bx \mid S_t(0)=S_t(1)=1) \log \frac{p(\bx \mid S_t(0)=S_t(1)=1)}{p(\bx)}d\bx
    \\
    &= \E\left[
  \frac{1-\rho(\bX,t)}{1-\bar\rho_t}
  \log\left\{
  w_0(\bX,t)
  \frac{1-\rho(\bX,t)}{1-\bar\rho_t}
  \right\}
  \middle|S_t(0)=1
  \right] \\
  &=
  \E\left[
  \frac{1-\rho(\bX,t)}{1-\bar\rho_t}
  \log w_0(\bX,t)
  \middle|S_t(0)=1
  \right] + \E\left[
  \frac{1-\rho(\bX,t)}{1-\bar\rho_t}
  \log\frac{1-\rho(\bX,t)}{1-\bar\rho_t}
  \middle|S_t(0)=1
  \right],
\end{align*}
we have that
\begin{align*}
    \psi_t - \psi_t^\text{mono} &= 
    \E\left[
  \left\{
  \frac{1-\rho(\bX,t)}{1-\bar\rho_t}-1
  \right\}
  \log w_0(\bX,t)
  \middle|S_t(0)=1
  \right] +
  \E\left[
  \frac{1-\rho(\bX,t)}{1-\bar\rho_t}
  \log\frac{1-\rho(\bX,t)}{1-\bar\rho_t}
  \middle|S_t(0)=1
  \right] \\[1em]
  &=
  -\frac{
  \Cov\{\rho(\bX,t),\log w_0(\bX,t)\mid S_t(0)=1\}
  }{
  1-\bar\rho_t
  }
  +
  \text{KL}\left\{
  p\{\bX\mid S_t(0)=S_t(1)=1\}
  \middle\|
  p\{\bX\mid S_t(0)=1\}
  \right\} \\
  &= \text{A} + \text{B}.
\end{align*}
Analogous to Proposition \ref{prop:sens-bias}, for term A notice Cauchy Schwarz implies
\[
\frac{
  | \Cov\{\rho(\bX,t),\log w_0(\bX,t)\mid S_t(0)=1\}|
  }{
  1-\bar\rho_t
  } \leq \frac{\sigma_{\rho_t}}{1-\bar\rho_t}\text{SD}\{ \log w_0(\bX,t) \mid S_t(0)=1\}.
\]
For term B, letting $\chi^2(P || Q)$ denote the Chi-square divergence between generic $P$ and $Q$ notice
\[
 \text{KL}\left\{
  p\{\bX\mid S_t(0)=S_t(1)=1\}
  \middle\|
  p\{\bX\mid S_t(0)=1\}
  \right\} \leq \log\left\{1 +  \chi^2\left\{
  p\{\bX\mid S_t(0)=S_t(1)=1\}
  \middle\|
  p\{\bX\mid S_t(0)=1\}
  \right\} \right\}
\]
by Jensen's inequality. Then, notice 
\begin{align*}
 & \ \ \chi^2\left\{
  p\{\bX\mid S_t(0)=S_t(1)=1\}
  \middle\|
  p\{\bX\mid S_t(0)=1\}
  \right\}\\  &= \E \left[ \left(\frac{p\{\bX\mid S_t(0)=S_t(1)=1\}}{ p\{\bX\mid S_t(0)=1\}} -1 \right)^2  \bigg| S_t(0)=1 \right]  \\
  &=\E\left[ \left(\frac{1-\rho(\bX,t)}{1-\bar \rho_t} -1\right)^2 \bigg| S_t(0)=1  \right] \\
  &=
  \frac{\E\left[ 
  \{\rho(\bX,t) - \bar \rho_t\}^2 \mid S_t(0)=1
  \right]}{(1-\bar \rho_t)^2} \\
  &= \frac{\sigma^2_{\rho_t}}{(1-\bar \rho_t)^2}.
\end{align*}
Recalling the form of $\psi_t - \psi_t^\text{mono}$ from above, it follows that
\begin{align*}
   \psi_t - \psi_t^\text{mono} &\leq   -\frac{
  \Cov\{\rho(\bX,t),\log w_0(\bX,t)\mid S_t(0)=1\}
  }{
  1-\bar\rho_t
  } + \log\left\{ 1 + \frac{\sigma^2_{\rho_t}}{(1-\bar \rho_t)^2}\right\}.
\end{align*}
Since the second term is non-negative, we have
\begin{align*}
    \psi_t  \in &\bigg[ \psi_t^\text{mono} -  \frac{\sigma_{\rho_t}}{1-\bar\rho_t}\text{SD}\{ \log w_0(\bX,t) \mid S_t(0)=1\}, \\
    & \ \  \psi_t^\text{mono} + \frac{\sigma_{\rho_t}}{1-\bar\rho_t}\text{SD}\{ \log w_0(\bX,t) \mid S_t(0)=1\} +  \log\left\{ 1 + \frac{\sigma^2_{\rho_t}}{(1-\bar \rho_t)^2}\right\} \bigg],
\end{align*}
recalling the earlier bound for $\Cov\{\rho(\bX,t),\log w_0(\bX,t)\mid S_t(0)=1\}$.

\paragraph{Estimation} One can construct estimators for the above upper and lower bounds, recalling that $\hat \psi_t^\text{mono}$ can be estimated with the methods from Section \ref{sec:methods}. Noting that 
\[
\text{SD}\{ \log w_0(\bX,t) \mid S_t(0)=1\} = \sqrt{\E[w_0(\bX,t)\{\log w_0(\bX,t)\}^2] - (\psi_t^\text{mono})^2},
\]
one can analogously construct a one-step estimator for $\text{SD}\{ \log w_0(\bX,t) \mid S_t(0)=1\}$ by constructing one-step estimators of the two terms above.  In particular, Supplementary Material \ref{app:alt-divergences} below provides methods for estimating $\E[w_0(\bX,t)\{\log w_0(\bX,t)\}^2]$, by noting it's a special case of the general functional $\E[h(w_0(\bX,t))]$ with $h(w) = w \{\log w\}^2$.

\section{Alternative divergence measures}
\label{app:alt-divergences}

The debiased estimation framework developed in Sections \ref{sec:identification}--\ref{sec:methods} extends more broadly to scalar divergences other than the KL divergence. Two  alternatives that have well-developed empirical roles in the divergence-estimation literature are the chi-square divergence $\chi^2_t$ and the squared Hellinger distance $H_t$
\begin{align*}
  \chi^2_t &= \int \frac{\{p(\bX|S_t(0)=S_t(1)=1) - p(\bX)\}^2}{p(\bX)} d\bx, \\
  H_t &= \frac{1}{2} \int \left( \sqrt{p(\bX|S_t(0)=S_t(1)=1)} - \sqrt{p(\bX)} \right)^2 d\bx,
\end{align*}
which under Assumptions \ref{as:consistency}-\ref{as:monotonicity} are identified by
\begin{align*}
\chi^2_t &= \E\left[(w_0(\bX,t) - 1)^2\right], &
H_t   &= \tfrac{1}{2}\E\left[\left(\sqrt{w_0(\bX,t)} - 1\right)^2\right].
\end{align*}
Notably, the identifying functionals for $\chi_t^2$, $H_t$ and the KL divergence $\psi_t$ can all be viewed as nonlinear functionals of the form $D_{h,t}=\E[h(w_0(\bX,t))]$, where $h(\cdot)$ is a nonlinear transformation of $w_0(\bX,t)$, with $h(w_0(\bX,t)) = (w_0(\bX,t) - 1)^2$ for $\chi_t^2$, $h(w_0(\bX,t) = \frac{1}{2}\left(\sqrt{w_0(\bX,t)} - 1\right)^2$ for $H_t$, and $h(w_0(\bX,t)) = w_0(\bX,t) \log w_0(\bX,t)$ for $\psi_t$.  Using analogous arguments to the ones used in Theorem \ref{thm:eif-thm},  for general functionals of this form we have
\begin{align*}
\varphi_{D_{h,t}}(\bm O)
=&\frac{\II(A=0)}{1-e(\bX)} 
\frac{h'(w_0(\bX,t))-C_{D_{h,t}}}{\mu_{0t}}
\{S_t-g_0(\bX,t)\}\\
&+h\{w_0(\bX,t)\}-C_{D_{h,t}}w_0(\bX,t)+C_{D_{h,t}}-D_{h,t},
\end{align*}
where
\[
C_{D_{h,t}}=\E\left[w_0(\bX,t)h'\{w_0(\bX,t)\}\right].
\]
For $h(w)=(w-1)^2$ and $h(w)=\tfrac12(\sqrt w-1)^2$, respectively, $C_{D_{h,t}}=2\chi_t^2$ and $C_{D_{h,t}}=H_t/2$, implying for these example divergence functionals
\[
\begin{aligned}
\varphi_{\chi_t^2}(\bm O)
=&\frac{\II(A=0)}{1-e(\bX)}
\frac{2\{w_0(\bX,t)-1\}-2\chi_t^2}{\mu_{0t}}
\{S_t-g_0(\bX,t)\} +\{w_0(\bX,t)-1\}^2-2\chi_t^2w_0(\bX,t)+\chi_t^2,\\
\varphi_{H_t}(\bm O)
&=\frac{\II(A=0)}{1-e(\bX)}
\frac{\tfrac12\{1-w_0(\bX,t)^{-1/2}\}-H_t/2}{\mu_{0t}}
\{S_t-g_0(\bX,t)\}+\tfrac12\{\sqrt{w_0(\bX,t)}-1\}^2
-\frac{H_t}{2}w_0(\bX,t)-\frac{H_t}{2}.
\end{aligned}
\] 

\clearpage 

\section{Additional Simulation Details}

\paragraph{Estimation} For our main simulation exercise, at each time point $t$, in the correctly-specified scenario we estimate $g_0(\bX,t)$ with a SuperLearner ensemble of a  main effects logistic regression with interactions; a ridge-penalized logistic regression with main effects, interactions and quadratic terms; and degree-two MARS. For the misspecified scenario we estimate only with a main effects logistic regression. Throughout, we use  the known treatment probability $e(\bX)=1/2$. We construct both one-step estimators through 5-fold cross-fitting \citep{kennedy2024semiparametric}.
\\\\
For our sensitivity analysis simulation, we follow the framework outlined in Appendix \ref{app:sens-bounds}. For simplicity we estimate $g_0(\bX,t)$ with a correctly specified logistic regression with main effects, interactions and quadratic terms.

\begin{table}[ht!]
    \centering
    \begin{tabular}{llr}
    \toprule
    Parameter & Role & Value \\ \midrule
    $\beta_0$ & baseline survival log-odds (intercept) & $5.70$ \\
    $\beta_1$ & decline in log-odds per unit $t/T$ & $4.00$ \\
    $(\alpha_1,\alpha_2,\alpha_3)$ & main covariate effects & $(1.05,0.46,0.35)$ \\
    $(\kappa_1,\kappa_2,\kappa_3)$ & quadratic effects & $(0.12,0.10,0.08)$ \\
    $(\gamma_{12},\gamma_{13},\gamma_{23})$ & pairwise interactions & $(0.50,0.50,0.50)$ \\
    $c$ & treated/control hazard ratio & $0.60$ \\
    $T$ & number of follow-up periods & $10$ \\ \midrule
    $\bX$ & baseline covariate distribution & $N(\boldsymbol 0,\mathbf I_3)$ \\
    $e(\bX)$ & known treatment propensity & $1/2$ \\
    \bottomrule
    \end{tabular}
    \caption{Data-generating parameters for the simulation study.}
    \label{tab:sim-params}
\end{table}

\clearpage

\section{Additional PRO-ACT Application Details}

\subsection{Analysis Data Construction}

\paragraph{ALSFRS-R.} We use the ALSFRS-R total only. Where it is missing but all twelve item-level fields are available, we reconstruct it as their sum.

\paragraph{Baseline windows.} Baseline ALSFRS-R is taken from the visit nearest enrollment within 14 days. Following \cite{beaulieu2021development}, baseline vital capacity (\% predicted) is taken as forced vital capacity where available, and slow vital capacity otherwise.

\paragraph{BMI.} BMI is computed  as $\text{weight}_{\text{kg}}/(\text{height}_{\text{m}})^2$ with unit-of-measure standardization across trials, clipping implausible values outside $[12, 80]$ to NA.

\paragraph{Pre-baseline progression rate.} Following \citet{westeneng2018prognosis}, $$\Delta\text{FRS} = (48 - \text{ALSFRS-R}_0)/\text{months from symptom onset},$$ set to \texttt{NA} when disease duration is missing or non-positive.

\paragraph{Site of onset.} Site of onset is coded as a binary indicator of bulbar onset.

\paragraph{Survival availability.} For each $t \in \{3, 6, 9, 12\}$ months, $S_t = 0$ if death occurred by $t \cdot 30$ days and $S_t = 1$ if death occurred later or the subject is observed alive beyond $t \cdot 30$ days. Subjects whose status at $t$ is undetermined---dropout before $t$ with no recorded death---are treated as censored. We discuss extensions which additionally account for censoring in Supplementary Material \ref{app:censoring}.

  \begin{table}[ht!]
      \centering
      \small
      \begin{tabular}{p{0.22\linewidth}p{0.70\linewidth}}
      \toprule
      Variable & Description \\ \midrule
      \multicolumn{2}{l}{\textit{Treatment and follow-up variables}} \\[0.25em]
      Treatment, $A$
          & Treatment assignment recorded in PRO-ACT as Active or Placebo. Because trial identifiers are
          masked, Active includes several investigational treatments. \\
      Survival, $S_t$
          & Whether the participant was alive at month $t$, for $t\in\{3,6,9,12\}$. Survival is
          undetermined if no death was recorded and follow-up ended before month $t$. \\
      ALSFRS-R outcome, $Y_t$
          & ALSFRS-R total score recorded within $\pm30$ days of month $t$ among participants known to be
          alive. Scores range from 0 to 48, with higher scores indicating better function.  \\ 
    Censoring, $R_t$
      & Whether the participant's survival status at month $t$ could be determined from the available
      follow-up. Participants with follow-up ending before month $t$ and no recorded death have $R_t=0$. \\
\midrule
      \multicolumn{2}{l}{\textit{Baseline covariates}} \\[0.25em]
      Age
          & Age at enrollment, in years. \\
      Male
          & Whether the participant was recorded as male. \\
      ALSFRS-R
          & ALSFRS-R total score from the visit closest to enrollment within 14 days. \\
      Vital capacity
          & Percent-predicted vital capacity from the visit closest to enrollment within 30 days. Forced
          vital capacity is used when available and slow vital capacity otherwise. \\
      Riluzole
          & Riluzole use at baseline. \\
      BMI
          & Baseline body mass index in $\mathrm{kg}/\mathrm{m}^2$, calculated after standardizing weight
          and height units across trials. \\
      $\Delta$FRS
          & Rate of functional decline before enrollment, calculated as $(48-\text{baseline ALSFRS-R})/
          (\text{months from symptom onset})$. Larger values indicate faster decline. \\
      Bulbar onset
          & Whether bulbar onset was recorded on the ALS history form. Participants with no recorded onset
          site were coded as non-bulbar. \\
      \bottomrule
      \end{tabular}
      \caption{Variables used in the PRO-ACT application.}
      \label{tab:proact-variables}
  \end{table}

\begin{table}[htbp]
  \centering
  \label{tab:as-covariate-means}
  \begin{tabular}{lrrrrr}
  \toprule
  Covariate
    & Baseline
    & 3 months
    & 6 months
    & 9 months
    & 12 months \\
  \midrule
  Age, years
    & 57.4 & 57.4 & 57.2 & 56.9 & 56.0 \\
  Male, \%
    & 63.6 & 63.4 & 63.1 & 63.2 & 61.6 \\
  ALSFRS-R
    & 37.54 & 37.57 & 37.67 & 37.83 & 38.21 \\
  Riluzole use, \%
    & 59.7 & 60.0 & 60.1 & 60.1 & 62.6 \\
  $\Delta$FRS, points/month
    & 0.632 & 0.629 & 0.618 & 0.602 & 0.581 \\
  Vital capacity, \% predicted
    & 82.0 & 82.1 & 82.2 & 83.1 & 85.1 \\
  BMI
    & 26.28 & 26.30 & 26.27 & 26.38 & 26.49 \\
  Bulbar onset, \%
    & 14.3 & 14.4 & 13.9 & 14.3 & 13.8 \\
  \bottomrule
  \end{tabular}
  \caption{Estimated baseline and time-$t$ survivor covariate means over follow-up.}
  \end{table}

\subsection{Assumptions}

We briefly discuss the plausibility of Assumptions \ref{as:consistency}-\ref{as:monotonicity} in our application. Consistency is only required among  participants receiving placebo, which is likely to hold.  Letting $\bX$ collect all baseline covariates outlined in Table \ref{tab:proact-variables}, Assumption \ref{as:positivity} similarly is likely to hold since the majority of trials contributing to PRO-ACT use 50/50 or 75/25 randomization, meaning no covariate profiles should be precluded from receiving active treatment or placebo. Since trial identifiers aren't provided by PRO-ACT, and since treatment allocation and covariate distributions can feasibly differ across trials, Assumption \ref{as:surv-unc} requires that $\bX$ sufficiently captures any factors differing across trials associated with both treatment assignment and survival. Since few past ALS trials have provided evidence of harmful effects of treatment, Assumption \ref{as:monotonicity} likely holds approximately. However, there do exist previous trials suggesting possible harmful treatment effects \citep{meininger2006pentoxifylline,gordon2007efficacy}.  To account for the possibility that a subset of these trials appear in our PRO-ACT sensitivity analysis in Section \ref{sec:application}, our sensitivity analysis explores modest violations to this Assumption.
\\\\
Though SACE estimation is not of direct interest in our application, we additionally discuss Assumptions \ref{as:out-unc} and \ref{as:cross-ex}. Like Assumption \ref{as:surv-unc}, Assumption \ref{as:out-unc} similarly requires $\bX$ to sufficiently capture prognostic factors of ALSFRS-R that differ across trials contributing to PRO-ACT. Assumption \ref{as:cross-ex} is a strong cross-world condition that additionally requires $\bX$ to contain sufficient prognostic information for the functional outcome ALSFRS-R, including within any single contributing trial rather than only across trials. Though $\bX$ does contain strongly prognostic features like baseline progression rate and vital capacity, it does not include other complementary prognostic factors like neurofilament light chain. We provide details on estimation of the SACE in Section \ref{app:censoring}.

\section{Accounting for Censoring Due to Dropout}
\label{app:censoring}

In this Section, we discuss extensions to our proposed methods that accommodate right censoring, or study dropout prior to death. We note that, since each divergence measure at a fixed time $t$ is a separate estimand depending only on the single conditional survival probability $g_0(\bX,t) = \PP(S_t=1 \mid A=0, \bX)$, censoring at time $t$ can be treated as a missing binary outcome problem, which we can similarly accommodate without imposing longitudinal structure across time points. Let $R_t$ be a censoring variable indicating whether a subject's survival status at $t$ is known, and let $\pi_t(\bX) = \PP(R_t=1 \mid A=0, \bX)$.

\begin{assumption}[Conditionally independent censoring]
\label{as:car}
$R_t \indep S_t \mid A, \bX$ for each $t$.
\end{assumption}
\begin{assumption}[Positivity of complete follow-up]
\label{as:cens-pos} For each $t$,
$\pi_t(\bX) > 0$ almost surely for all $\bX$ with $\PP(S_t(0)=S_t(1)=1 \mid \bX) > 0$.
\end{assumption}
Assumption \ref{as:car} requires that measured baseline covariates and treatment explain all systematic factors influencing both dropout and survival by time $t$. For each $t$, Assumption \ref{as:cens-pos} requires all possible survivor covariate profiles to have a positive probability of complete follow-up up to time $t$. 

Under these additional Assumptions, $\theta_{kt}$ and $\psi_t$ are identified by the functionals
\begin{align*}
    \tilde \theta_{kt} &= \frac{\E[X_k \cdot \tilde g_0(\bX,t)]}{\E[\tilde g_0(\bX,t)]} = \E[X_k \cdot \tilde w_0(\bX,t)] \\
    \tilde \psi_t &= \E[\tilde w_0(\bX,t) \log \tilde w_0(\bX,t)],
\end{align*}
where $\tilde g_0(\bX,t) := \PP(S_t = 1|\bX, A=0, R_t=1)$ and $\tilde w_0(\bX,t) = \tilde g_0(\bX,t)/\E[\tilde g_0(\bX,t)]$.
The corresponding EIFs are nearly identical to those for $\theta_{kt}$ and $\psi_t$:
\begin{align*}
 & \varphi_{\tilde\theta_{kt}}(\bm O)
=
\frac{1}{\mu_{0t}}
\left[
\frac{\mathbb{I}(A=0)R_t}{\{1-e(\bX)\}\pi_t(\bX)} \{S_t - \tilde g_0(\bX,t)\}(X_k - \theta_{k,t})
+
\tilde g_0(\bX,t)(X_k - \theta_{k,t})
\right],   \ \ \ \text{and} \\
&\varphi_{\tilde \psi_t}(\bm O) = 
\frac{\mathbb{I}(A=0)R_t}{\{1-e(\bX)\}\pi_t(\bX)}
\frac{\log \tilde w_{0}(\bX,t)-\psi_t}{\mu_{0t}}
\bigl(S_t-\tilde g_0(\bX,t)\bigr)
+
\tilde w_0(\bX,t)\bigl(\log \tilde w_0(\bX,t)-\psi_t-1\bigr)
+1.
\end{align*}
One can then construct plug-in and one-step estimators using the same strategy outlined in Section \ref{sec:methods}.

\subsection{SACE Estimation}

Analogous to our estimator for $\theta_{kt}$, in our PRO-ACT application in Section \ref{sec:application} we estimate the SACE $\tau_t$ using a ratio one-step estimator, adapting the approach from \citep{tchetgen2014identification}. Since $Y_t$ can be missing regardless of whether $R_t=1$, we let $Q_t$ be an indicator of whether the functional outcome is observed at time $t$. Then, under Assumptions \ref{as:consistency}-\ref{as:cens-pos}, and the additional condition  $Q_t \indep Y_t \mid A,\bX,S_t=1$, $\tau_t$ is identified by
\[
\tau_t =
\frac{\E\left[g_0(\bX,t)\left\{m_1(\bX,t)-m_0(\bX,t)\right\}\right]}{\E[g_0(\bX,t)]
},
\]
where $m_a(\bX,t) = \E[Y_t \mid A=a, \bX, Q_t=1]$.
Then, a one-step estimator for $\tau_t$ is
$
\hat\tau_t^{\mathrm{OS}}=\frac{\hat N_t^{\mathrm{OS}}}{\hat D_t^{\mathrm{OS}}},
$
where 
\begin{align*}
  \hat N_t^{\mathrm{OS}} &= \sumn \Bigg[\left\{\hat g_0(\bX_i,t)+\frac{\II(A_i=0)R_{ti}\{S_{ti}-\hat g_0(\bX_i,t)\}}{\{1-\hat
  e(\bX_i)\}\hat\pi_t(\bX_i)}\right\}\{\hat m_1(\bX_i,t)-\hat m_0(\bX_i,t)\} \\
  &\qquad+\hat g_0(\bX_i,t)\left\{\frac{\II(A_i=1)Q_{ti}\{Y_{ti}-\hat m_1(\bX_i,t)\}}{\hat e(\bX_i)\hat
  q_1(\bX_i,t)}-\frac{\II(A_i=0)Q_{ti}\{Y_{ti}-\hat m_0(\bX_i,t)\}}{\{1-\hat e(\bX_i)\}\hat q_0(\bX_i,t)}\right\}\Bigg], \\
  \hat D_t^{\mathrm{OS}} &= \sumn\left[\hat g_0(\bX_i,t)+\frac{\II(A_i=0)R_{ti}\{S_{ti}-\hat g_0(\bX_i,t)\}}{\{1-\hat
  e(\bX_i)\}\hat\pi_t(\bX_i)}\right],
\end{align*}
and $q_a(\bX,t)=\PP(Q_t=1\mid A=a,\bX)$.

\end{document}

%% file: preamble.tex
\usepackage[utf8]{inputenc}

\usepackage{geometry}
\usepackage{natbib}
\usepackage{amsmath}
\usepackage{amssymb}
\usepackage{xcolor}
\usepackage{hyperref}
\usepackage{booktabs}
\usepackage{amsthm}
\usepackage{bm}
\usepackage{pdflscape}
\usepackage{setspace}
\usepackage{graphicx}
\usepackage{subcaption}
\usepackage{soul}
\usepackage{placeins}

\allowdisplaybreaks

\hypersetup{colorlinks=true,linkcolor=blue,filecolor=blue,citecolor=blue}

\usepackage{tikz}
\usepackage[labelfont=bf]{caption}
\usepackage{relsize}
\usepackage{xifthen}
\usepackage{amsmath,amssymb,xcolor}
\usepackage{pgf,tikz}
\usetikzlibrary{arrows,shapes.arrows,shapes.geometric,
shapes.multipart,decorations.pathmorphing,positioning,
swigs}

\newtheorem{theorem}{Theorem}
\newtheorem{assumption}{Assumption}

\numberwithin{intassumption}{assumption}

\newcommand{\E}{\mathbb{E}}
\newcommand{\PP}{\mathbb{P}}
\newcommand{\bX}{\boldsymbol X}

\newcommand{\bx}{\boldsymbol x}
\newcommand{\II}{\mathbb{I}}
\newcommand{\indep}{\perp \!\!\! \perp}

\newcommand{\eqdef}{\overset{\text{def}}{=}}

\newcommand{\Cov}{\text{Cov}}

\newcommand{\sumn}{\frac{1}{n}\sum_{i=1}^n}

\DeclareMathOperator{\expit}{expit}

\newtheorem{proposition}{Proposition}